\documentclass[11pt]{article}
\usepackage{amsmath, amsthm, amssymb, amsthm, dcolumn}

\usepackage{graphicx} 
\usepackage{color}
\usepackage{soul}
\usepackage{float}
\usepackage[utf8]{inputenc}
\usepackage{subcaption}
\usepackage[titletoc, title]{appendix}
\usepackage{natbib}
\usepackage{tabu,multirow}
\usepackage[titletoc]{appendix}
\usepackage{import}
\usepackage{graphicx}
\usepackage{ragged2e}

\usepackage[utf8]{inputenc}
\usepackage[flushleft]{threeparttable}
\usepackage{dcolumn}
\usepackage{caption}
\usepackage{soul}
\usepackage[table]{xcolor}

\graphicspath{ {Pictures/} }

\setcitestyle{authoryear, open={((},close={)}}

\begin{document}

\begin{titlepage}
	\begin{center}
	{\huge A short term credibility index for central banks under inflation targeting: an application to Brazil\par }
	 \vspace{1cm}
	 \renewcommand{\thefootnote}{\alph{footnote}}
	{\Large Alain Hecq\footnote{Maastricht University School of Business and Economics}, João Victor Issler\footnote{FGV EPGE Brazilian School of Economics and Finance} and Elisa Voisin\footnotemark[1]\footnote{Corresponding author : Elisa Voisin, Maastricht University, Department of Quantitative Economics, School of Business and Economics, P.O.box 616, 6200 MD, Maastricht, The Netherlands. Email: e.voisin@maastrichtuniversity.nl.\par}
	} \\
	\vspace{0.2cm}

	\vspace{1cm}
	{\large May 2022 \par}
	
	\vspace{1cm}
	\end{center}
    \begin{abstract}
        This paper uses predictive densities obtained via mixed causal-noncausal autoregressive models to evaluate the statistical sustainability of Brazilian inflation targeting system with the tolerance bounds. The probabilities give an indication of the short-term credibility of the targeting system without requiring modelling people's beliefs. We employ receiver operating characteristic curves to determine the optimal probability threshold from which the bank is predicted to be credible. We also investigate the added value of including experts predictions of key macroeconomic variables.
    \end{abstract}

	\vspace{0.5cm}
	
\end{titlepage}

\graphicspath{ {Pictures/} }

\section{Introduction}\label{bra-sec:Intro}
\begin{figure}[h!]
\centering
\includegraphics[width=\textwidth,trim={0cm 1cm 0cm 1cm},clip]{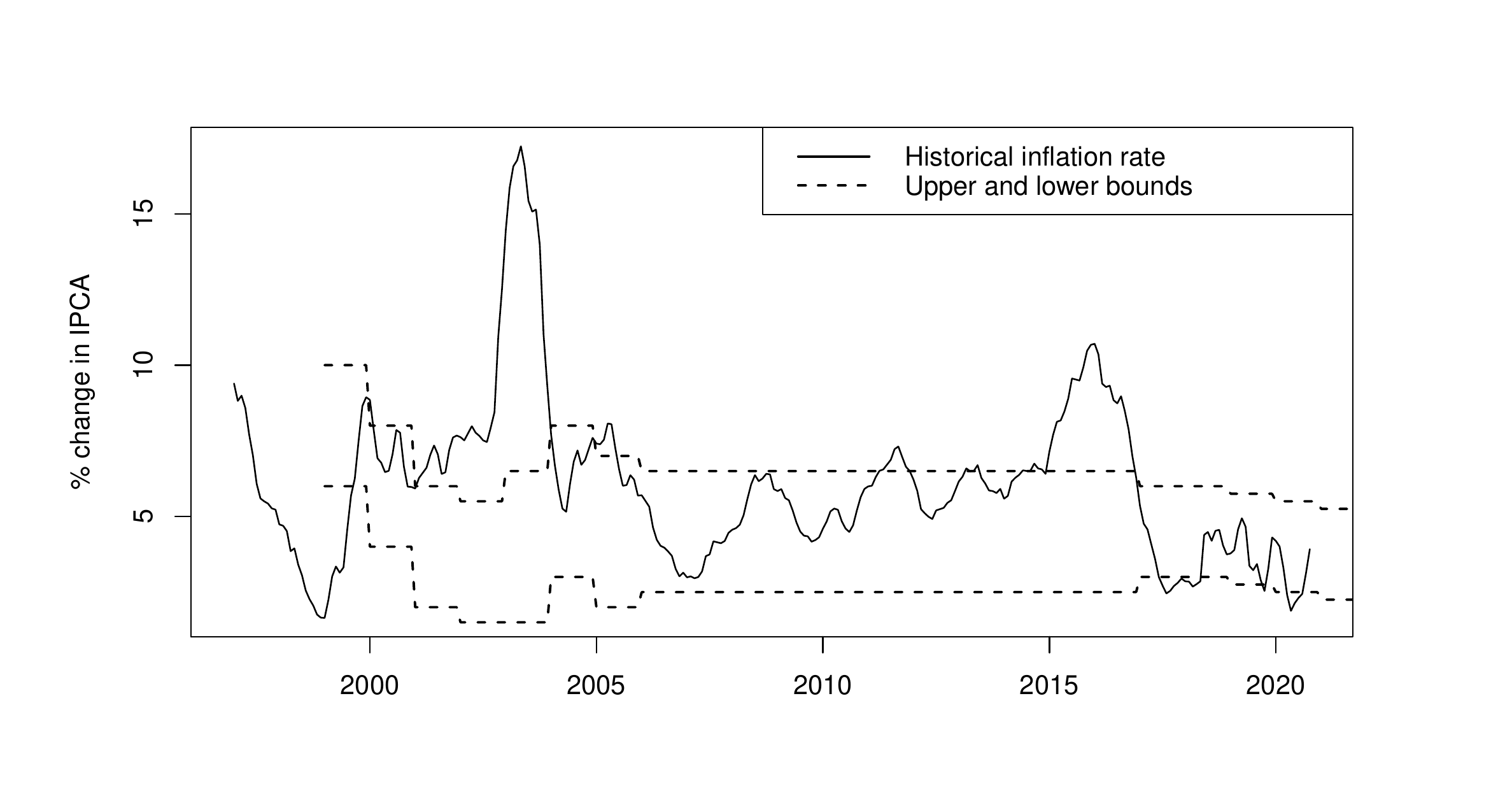} 
\caption{Annual inflation rate in Brazil and the target bounds}
\label{bra-fig:inflation_rate}
\end{figure}

With the introduction of the Real Plan in July 1994,
and later in 1999 with the \textit{Inflation Targeting Regime}, actual Brazilian inflation has been substantially smaller compared to the previous
hyperinflation period of the 1970s and 80s. Annual inflation computed with
the Extended National Consumer Price Index (IPCA\footnote{%
The IPCA targets population families with household income ranging from 1 to
40 minimum wages. This income range guarantees a 90\% coverage of families
living in 13 geographic zones: metropolitan areas of Bel\'{e}m, Fortaleza,
Recife, Salvador, Belo Horizonte, Vit\'{o}ria, Rio de Janeiro, S\~{a}o
Paulo, Curitiba, Porto Alegre, as well as the Federal District and the
cities of Goi\^{a}nia and Campo Grande. Basket items include Food and
Beverages, Housing, Household Articles, Wearing Apparel, Transportation,
Health and Personal Care, Personal Expenses, Education and Communication.}),
the reference variable for the Brazilian \textit{Inflation-Targeting Regime}, is
6.07\% on average between January 1997 and October 2020. Figure \ref{bra-fig:inflation_rate} displays the IPCA annual inflation rate for that period, together with the
upper and lower tolerance bounds set within the regime. These tolerance bounds were 2 to 2.5 percentage points from the target inflation until 2017 and are now 1.5 percentage points above and below the target. With the noticeable exception of years 2001-2003 and 2015-2016, the Central Bank of Brazil (BCB) succeeded to ensure that the IPCA's annual
inflation remains within or close to the tolerance interval. \\

The question that we want to answer in this paper is whether, in any given month, it is sustainable for inflation to stay within the actual tolerance bounds set in advance by the \textit{Inflation-Targeting Regime} of the Brazilian Central Bank (BCB). We do this by computing the conditional probability that actual inflation stays within the bounds in the near future (1- to 6-month ahead) using a forward-looking approach that allows current inflation to depend on future inflation as well on lagged inflation. This is the first original contribution of this paper. \\

Applying the techniques discussed in this paper to other countries is straightforward, as long as we are able to obtain their respective  targets and bounds. So, the second original contribution of this paper is to propose a new methodological approach to measure the credibility of central banks under an \textit{Inflation-Targeting Regime}. Due to the nature of the problem we were set to answer, this methodology is related to short-term credibility as opposed to longer term credibility indices that have been proposed in the literature. Hence, we provide a complement to these techniques. \\

Although we employ a forward-looking approach, the information used to compute these probabilities across time only condition on current information. Indeed, we use a mixed causal-noncausal model -- MAR($r,s)$) -- a process with a \textit{lag} polynomial of order $r$ (as in the usual autoregressive AR($r)$ model), but with a \textit{lead} polynomial of order $s$ as well. The introduction of a forward component makes sense for modelling inflation, since many economic models link current inflation to their future expected value which is captured by the forward component introduced in the MAR($r,s)$ class. Moreover, this class also allows modelling parsimoniously nonlinear features in inflation, which the simpler linear setup misses. MAR models were introduced by \citet{lanne2012optimal} and refined by \citet{filtering}. Here we use the framework proposed in \citet{voisin2019forecasting,oilpaper}. \\

We do showcase the Brazilian experience, despite the fact that the techniques employed in this paper have a broader application. Inflation has been a major problem for this country over the years and inflation has been in and out the bounds according to Figure \ref{bra-fig:inflation_rate}. So, we would have two polar cases at hand: when the central bank policy had the desired effect on inflation and when it had not. Another interesting feature about Brazil is that, as far as we know, the BCB has the most complete data base on inflation expectations on earth -- The \textit{Focus} data base. It contains high frequency (daily) data on inflation expectations (also on other key macroeconomic variables) from current month all the way to 12 months ahead, and then on yearly intervals. This allows expectations data to be used to help forecast future inflation as a proxy for it, employing the  MARX model proposed by \citet{hecq2020mixed}, where strictly-exogeneous regressors are used in the model. \\

Most of the literature on central-bank credibility measures whether people expect or not that the central bank will meet their target; see \citet{blinder2000central}. This implies the need for modelling people's expectations regarding future inflation. As shown by \citet{issler2019central}, and the references therein, it is not straightforward to compute people's beliefs, and results can change substantially based on the method being employed. This paper instead focuses on the actual inflation process and asks whether it will remain within the announced bounds. It does not require modelling people's beliefs. It is based on the econometric properties and the dynamics of inflation using a forward-looking approach. \\

Empirically, the paper first identifies the MAR$(r,s)$ model on IPCA annual inflation. We found that the 12-month inflation rate follows a MAR(1,1) process with the error term having a Student's $t$ distribution with 3.25 degrees of freedom. This is our benchmark case, but we also experiment with alternative specifications. Using the approach in %
\citet{voisin2019forecasting,oilpaper}, and our benchmark specification, we evaluate the probabilities that the inflation rate stays on track in the future within the announced target bounds. \\

Finally, in our last original contribution, we compare the empirical results of our proposed short-term index to those of the literature by employing a receiver operating characteristic curve (ROC curve), something that is not common in the literature. For every measure of credibility, the ROC curve plots the true-positive and the false-positive rates for different probability thresholds used in classifying states (credible vs. non-credible). Our results show promise for the method propose in this paper as a complement to the existing ones, which are focused on credibility at longer horizons. \\

The rest of the paper is as follows. Section \ref{bra-sec:MAR} provides a summary of the model and estimation methods used here. Section \ref{bra-sec:Pred} summarizes the main methods that have been developed to forecast with MAR models. Section \ref{bra-sec:Application} provides the
estimated probabilities for the inflation to stay within announced bounds at various horizons. In Section \ref{bra-sec:MARX} we ask whether adding the information of experts forecasters from the \textit{Focus} database helps forecasting the conditional probabilities computed here. Section \ref{bra-sec:Credibility} compares our short-term credibility measure with existing credibility indices presented in the literature. Section \ref{bra-sec:Conclusion} concludes.
\section{Mixed causal-noncausal models}\label{bra-sec:MAR}

\subsection{Notation}

An MAR($r,s)$ process $y_t$ depends on its $r$
lags as for usual autoregressive processes but also on its $s$ leads in the
following multiplicative form%
\begin{equation*}
\Phi(L)\Psi(L^{-1})y_{t}=\varepsilon_{t},\label{bra-eq:MAR_definition}%
\end{equation*}
with $L$ is the backshift operator, i.e., $Ly_{t}=y_{t-1}$ gives lags and
$L^{-1}y_t=y_{t+1}$ produces leads. When $\Psi(L^{-1})=(1-\varphi
_{1}L^{-1}-...-\varphi_{s}L^{-s})=1,$ namely when $\varphi_{1}=...=\varphi
_{s}=0,$ the process $y_{t}$ is a purely causal autoregressive process,
denoted AR($r$,0) or simply AR($r$) model $\Phi(L)y_{t}=\varepsilon_{t}$.
The process is a purely noncausal AR($0,s)$ model $\Psi(L^{-1})y
_{t}=\varepsilon_{t},$ when $\phi_{1}=...=\phi_{r}=0$ in $\Phi(L)=(1-\phi
_{1}L-...-\phi_{r}L^{r}).$ The roots of both the causal and noncausal
polynomials are assumed to lie outside the unit circle, that is $\phi(z)=0$
and $\varphi(z)=0$ for $|z|>1$ respectively. These conditions imply that the
series $y_{t}$ admits a two-sided moving average (MA) representation $y
_{t}=\sum_{j=-\infty}^{\infty}\psi_{j}\varepsilon_{t-j},$ such that $\psi
_{j}=0$ for all $j<0$ implies a purely causal process $y_{t}$ (with respect to
$\varepsilon_{t}$) and a purely noncausal model when $\psi_{j}=0$ for all
$j>0$ \citep{lanne2011noncausal}. Error terms $\varepsilon_{t}$ are assumed \textit{i.i.d.} (and not only weak white noise) non-Gaussian to ensure the identifiability of the causal and the
noncausal part \citep{breid1991maximum}. \\

There is a increasing literature making use of MAR models; see among others \citet{karapanagiotidis2014dynamic}, \citet{hencic2015noncausal}, \citet{filtering}, \citet{lof2017noncausality}, \citet{hecq2019identification},
\citet{bec2020mixed}, \citet{gourieroux2020convolution}, \citet{gourieroux2021forecast}.

\subsection{Estimation results on IPCA}
Let $\pi_t$ denote the year-on-year inflation rate in Brazil at time \textit{t}. The hybrid New Keynesian Phillips Curve (NKPC) regression is such as%
\begin{equation}\label{bra-eq:NKPC}
\pi_{t}=\gamma_{f}\mathbb{E}_{t}\big[\pi_{t+1}\big]+\gamma_{b}\pi_{t-1}+\beta
x_{t}+\epsilon_{t},
\end{equation}
where $\mathbb{E}_{t}[\cdot]$ the conditional expectation at time $t$, $x_{t}$
is a measure for marginal costs, which is not directly observable (a potential proxy can be the output gap) and $\epsilon_{t}$ an
$i.i.d.$ error term. Adding and subtracting $\gamma_{f}\pi_{t+1}$ and
rearranging terms, gives
\begin{equation*}
\pi_{t}=\gamma_{f}\pi_{t+1}+\gamma_{b}\pi_{t-1}+\underbrace{\beta x_{t}%
+\gamma_{f}\left(  \mathbb{E}_{t}\big[\pi_{t+1}\big]-\pi_{t+1}\right)  +\epsilon_{t}%
}_{\equiv\eta_{t+1}},
\end{equation*}
where the newly defined disturbance term $\eta_{t+1}$ consists of three
different parts: ($i$) the expectation error $\left(  \mathbb{E}_{t}\big[\pi
_{t+1}\big]-\pi_{t+1}\right)  $ which is assumed $i.i.d.$ following the literature
on rational expectations models, ($ii$) the marginal costs variable $x_{t}$
and ($iii$) an \textit{i.i.d.} error $\epsilon_{t}$. Hence, the time series properties of $\eta_{t+1}$ will depend on those of $x_t$, but $x_t$ is assumed to be adequately approximated by a finite-order autoregression  \citep{lanne2013autoregression}. Subsequently, the newly obtained
equation is divided by $\gamma_{f}$ and lagged by one period to obtain
$(1-\gamma_{f}^{-1}L+\gamma_{f}^{-1}\gamma_{b}L^{2})\pi_{t}=-\gamma_{f}%
^{-1}\eta_{t}.$ Next, $a(z)\equiv(1-\gamma_{f}^{-1}z+\gamma_{f}^{-1}\gamma
_{b}z^{2})$ can be written as the product of two polynomials, i.e.,
$a(z)=(1-\phi z)(1-\varphi^{\ast}z)$ with $|\phi|<1$ and $|\varphi^{\ast}|>1$
for plausible values of $\gamma_{f}$ and $\gamma_b$ leading to a stable mixed
causal noncausal formulation. See \citet{lanne2013autoregression} for details. The MAR models we consider in this paper are not a direct mapping of the NKPC but rather an estimation of the dynamics emanating from the transformations of the Philips curve mentioned above. \\

We have used the MARX package developed by \citet{MARX}\footnote{We look
at different starting value to avoid the bimodality trap of the MAR(r,s)
estimated coefficients. This is not implemented in MARX so far. Note that 
\citet{hecq2020mixed} show how to estimate such a model directly
without solving for the exogenous variable $x_{t}.$ This will be estimated in Section \ref{bra-sec:MARX}.} and found on the whole
sample an MAR(1,1) with a Student's $t$ with 3.25 degrees of freedom. Standard
errors computed as in \citet{hecq2016identification} are in brackets.%
\begin{equation}\label{bra-eq:MAR_infl}
(1-\underset{(\textit{0.035})}{0.58}L)(1-\underset{(\textit{0.016})}{0.94}L^{-1})\pi_{t}=\varepsilon
_{t}, \qquad \varepsilon_{t}\sim t(3.25)
\end{equation}

\begin{figure}[h!]
\begin{subfigure}{\textwidth}
  \centering
  \includegraphics[width=\linewidth,trim={0cm 1cm 0cm 1cm},clip]{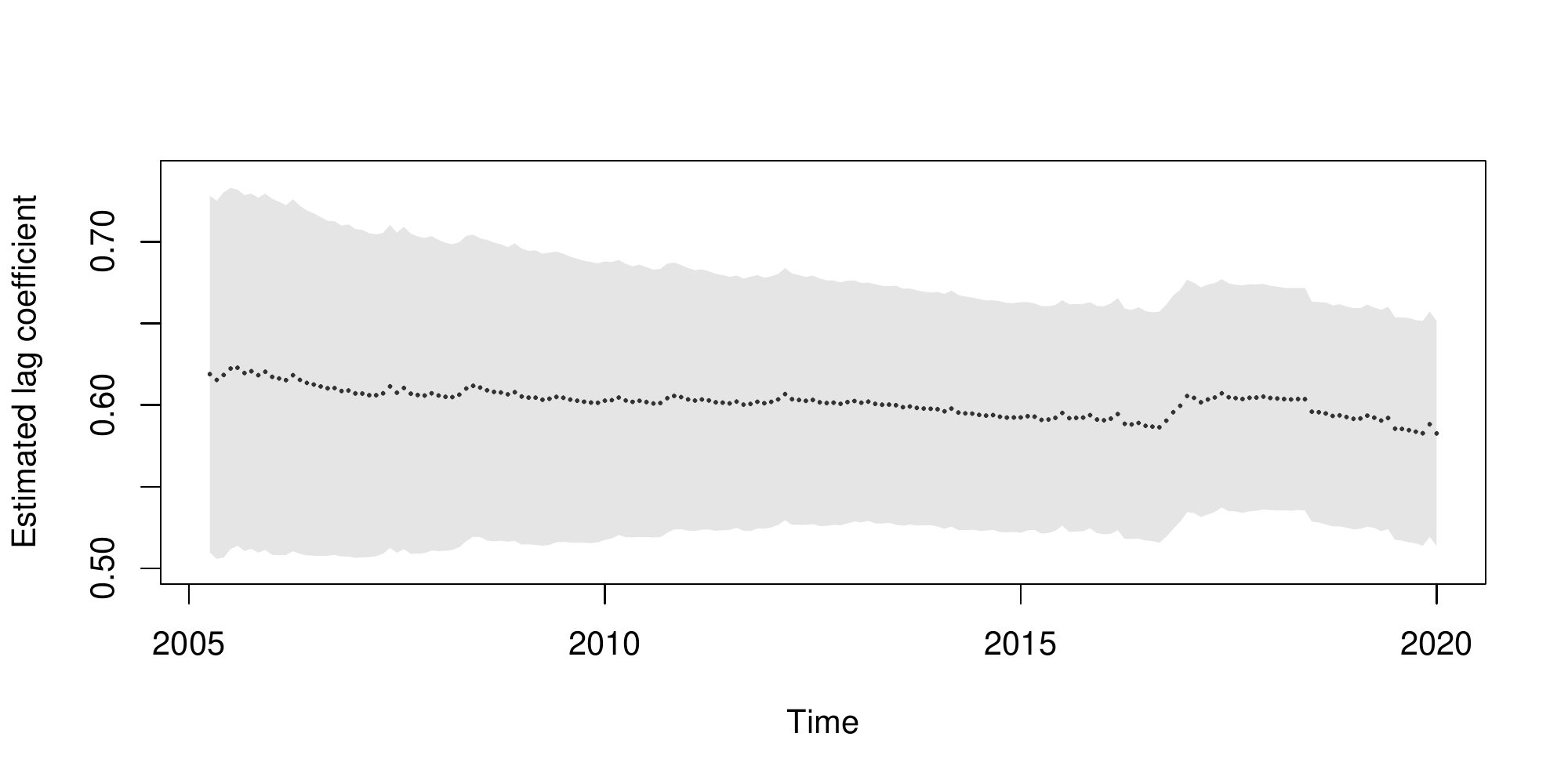}
\end{subfigure}%
\newline
\begin{subfigure}{\textwidth}
  \centering
  \includegraphics[width=\linewidth,trim={0cm 1cm 0cm 1.5cm},clip]{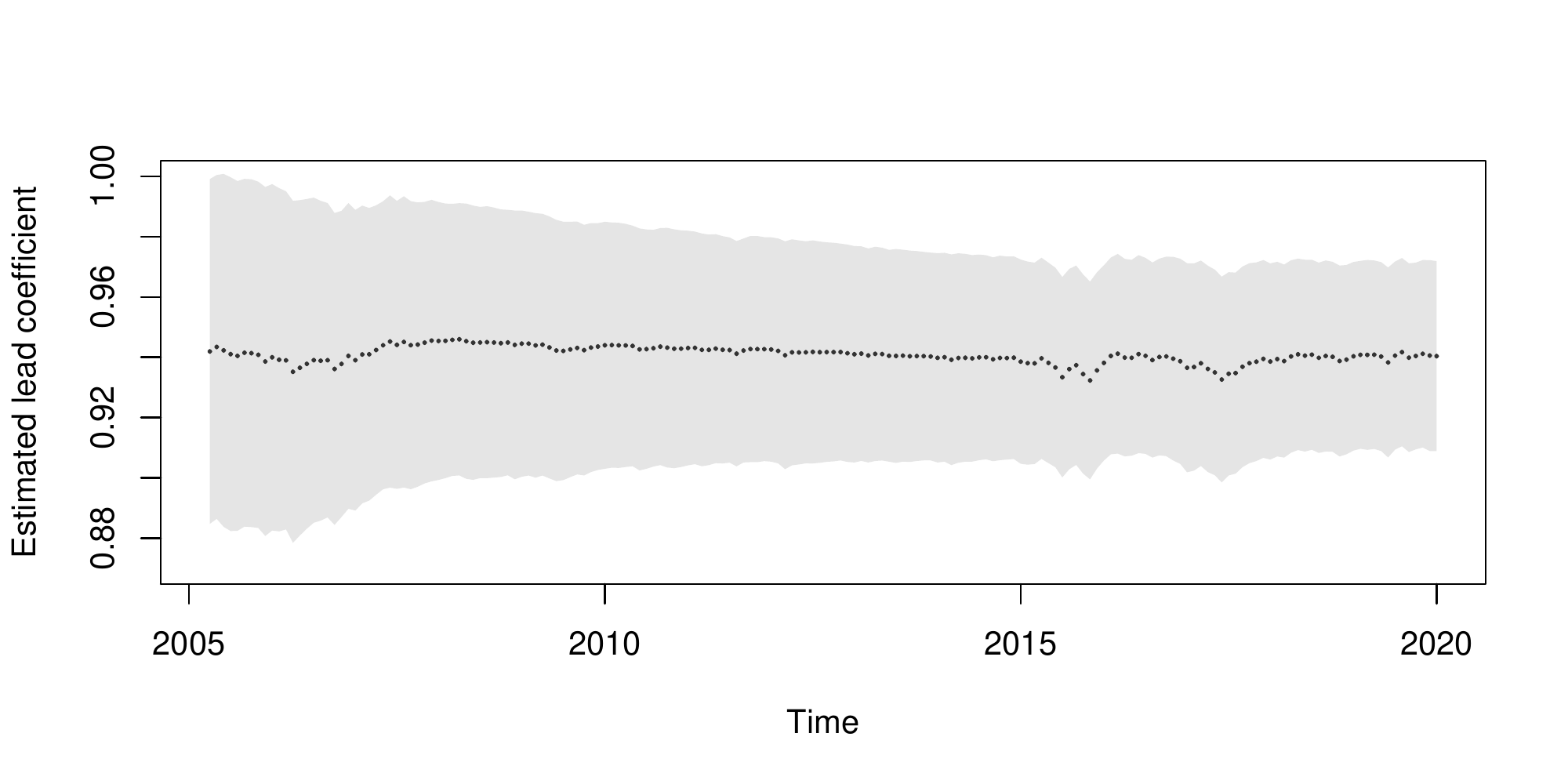}
\end{subfigure}%
\newline
\begin{subfigure}{\textwidth}
  \centering
  \includegraphics[width=\linewidth,trim={0cm 1cm 0cm 1.5cm},clip]{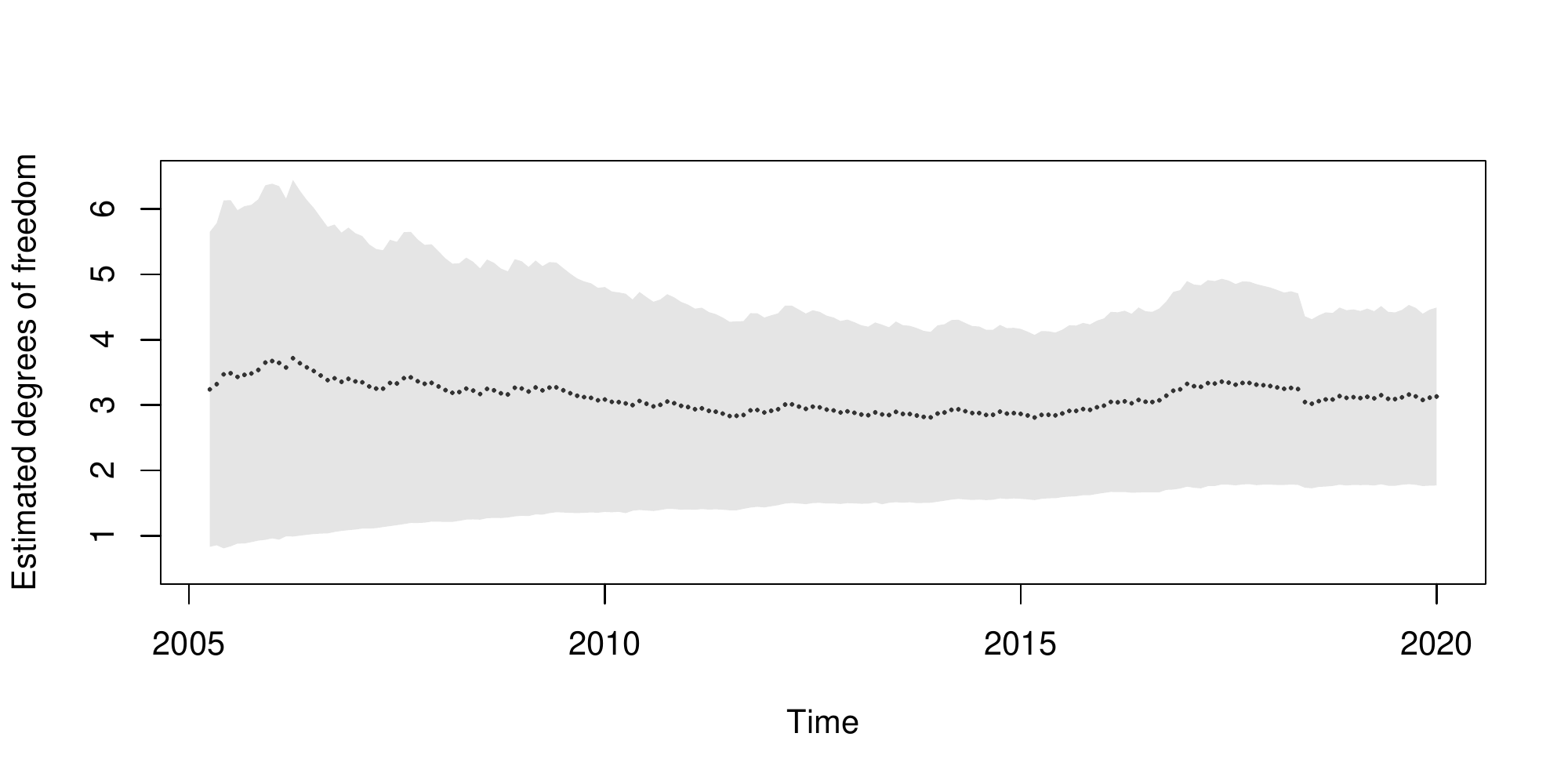}
\end{subfigure}
\vspace{-0.2cm}
\caption{Recursive estimates of the coefficients based on end point of sample and their 95\% C.I.}
\label{bra-fig:exp_window_estimation}
\end{figure}

To analyse the stability of the estimation we recursively estimate the orders of the MAR(\textit{r},\textit{s}) model and the corresponding coefficients with an expanding window. The initial sample goes from January 1997 to April 2005 (100 data points) and the last one goes to January 2020 (277 data points). Note that at each point the model identified was an MAR(1,1), we hence only provide the graphs of the recursive estimates of the lag coefficient, the lead coefficient and the degrees of freedom of the Student's \textit{t} distribution in Figure \ref{bra-fig:exp_window_estimation} with the corresponding 95\% confidence interval.\\

We can see that adding data points does not affect the value of the estimated coefficients. Over the 177 points added, the value of the lag coefficients slightly decreases from 0.62 to 0.58 whereas the lead coefficient varies between 0.936 and 0.948 and stabilises around 0.944 towards the end of the sample. The degrees of freedom vary between 3 and 3.7 and converges towards 3.25 towards the end of the sample. Hence, the slightly different models estimated in real time for recursive out of sample forecasts will not significantly affect the calculated probabilities and they will therefore be comparable. Note that from the correlogram of the residuals we might detect a seasonal MA(12) component. This feature is probably due to the construction of the year on year monthly inflation. Over differencing at some seasonal frequencies can introduce that moving average pattern. We keep working with this annual series as this is the target variable considered by the central bank. For our investigation that component mainly impacts standard errors. Mixed models with MA components (genuine or spurious) are also out of the scope of this paper.\\

Expanding the multiplicative model \eqref{bra-eq:MAR_infl} we obtain, 

\begin{equation*}
    \begin{split}
        \pi_t&= \frac{0.58}{1+0.58\times0.94}\pi_{t-1} + \frac{0.94}{1+0.58\times0.94}\pi_{t+1} + \frac{1}{1+0.58\times0.94}\varepsilon_{t} \\
        &= 0.38\pi_{t-1} + 0.61\pi_{t+1} + \varepsilon^*_{t}.
    \end{split}
\end{equation*}
The obtained coefficients for the lag and lead are of the same magnitude as in the literature on the NKPC, and we can notice that they add up to close to unity, which is often a restriction imposed for model identification \citep[see among others][]{gali2005robustness, nason2008identifying}.

\section{Predicting the probabilities to stay in the bounds}\label{bra-sec:Pred}

With Cauchy- and Levy-distributed errors, the conditional density of an MAR(\textit{r},\textit{s}) process admits closed-form expressions, this is however not the case for Student's \textit{t} distributed processes \citep{gourieroux2017local}. Assuming Student's \textit{t} errors offers more flexibility than the Cauchy distribution, which might be too extreme, especially for inflation series that are not particularly volatile. Hence, in the absence of closed-form expressions for the predictive density, two approximations methods have been developed. The first one is based on simulations and was proposed by \citet{lanne2012optimal}. The second one employs past realised values instead of simulations and was proposed by \citet{filtering}. However, as the latter becomes too computationally demanding when the forecast horizon increases, \citet{filtering} proposed a Sampling Importance Resampling (henceforth SIR) algorithm facilitating longer horizon forecasts with this method. While the algorithm does not always work for extreme events, it provides accurate results when the variables are rather stable, which is the case with the inflation series that we investigate here. Overall, both approaches use the decomposition of the mixed process into a causal and a noncausal component as such
\begin{equation}
    \begin{split}
        u_t &\equiv \Phi(L) \pi_t \\
        v_t &\equiv \Psi(L^{-1}) \pi_t.
    \end{split}
    \label{bra-eq:MAR.filter}
\end{equation}

The process $u_t$ is the purely noncausal component of the errors, on which we will focus. In this analysis, since the inflation series is an MAR(1,1), the process $u_t$ is a purely noncausal process of order 1,
\begin{equation}
    \begin{split}
        \Psi(L^{-1})u_t &= \varepsilon_t \\
        u_t&=\psi u_{t+1}+\varepsilon_t.
    \end{split}
    \label{bra-eq:MAR.filter_u}
\end{equation}

\subsection{Simulations-based approach}
The purely noncausal component of the errors, $u_t$, can be expressed as an infinite sum of future error terms in its MA representation. The stationarity of the process ensures the existence of an integer \textit{M} large enough to approximate this infinite representation as such \citep{lanne2012optimal}, 
\begin{equation}
    u_{t}\approx\sum_{i=0}^{M}\psi^i\varepsilon_{t+i},
    \label{bra-eq:u.in.eps}
\end{equation}

Let $lb_t$ and $ub_t$ be the lower and upper bound for inflation in Brazil assessed for time \textit{t}. We are interested in the conditional probabilities that inflation will be within the bounds at a given horizon \textit{h}, where \textit{T} is the last observed point in the sample,
\begin{equation}
\resizebox{0.93\textwidth}{!}{$
    \begin{aligned}
        \mathbb{P}\Big(lb_{T+h}\leq\pi_{T+h}^*\leq ub_{T+h}\big|\mathcal{F}_T\Big)&=\mathbb{P}\Big(\pi_{T+h}^*\leq ub_{T+h}\big|\mathcal{F}_T\Big)-\mathbb{P}\Big(\pi_{T+h}^*\leq lb_{T+h}\big|\mathcal{F}_T\Big) \\
        &=\mathbb{E}_T \bigg[\mathbf{1}\big(\pi^*_{T+h}\leq  ub_{T+h}\big)-\mathbf{1}\big(\pi^*_{T+h}\leq  lb_{T+h}\big)\bigg] 
    \end{aligned}
    $}
    \label{bra-eq:sim.estimator_1}
\end{equation} 

The indicator function $\mathbf{1}()$ is equal to 1 when the condition is met and 0 otherwise. \\

Since $\pi_t=\phi \pi_{t-1}+u_t$, by recursive substitution and using the approximation equation \eqref{bra-eq:u.in.eps}, we obtain,
\begin{equation}
    \begin{split}
        \pi_{T+h}&=\phi^h \pi_T+\sum_{i=0}^h \phi^i u_{T+h-i}\\
        &\approx\phi^h \pi_T+\sum_{i=0}^h\sum_{j=0}^{M-h-i}\phi^i\psi^j\varepsilon_{T+h-i+j},
    \end{split}
\end{equation}
where \textit{M} is the truncation parameter introduced in Equation \eqref{bra-eq:u.in.eps}. Substituting this approximation in \eqref{bra-eq:sim.estimator_1}, an approximation of the conditional probabilities is the following, 
\begin{equation}
\resizebox{0.93\textwidth}{!}{$
    \begin{aligned}
        \mathbb{P}\Big(lb_{T+h}\leq\pi_{T+h}^*\leq ub_{T+h}\big|\mathcal{F}_T\Big)\approx
        \mathbb{E}_T \Bigg[&\mathbf{1}\bigg( \pi_T+\sum_{i=0}^h\sum_{j=0}^{M-h-i}\phi^i\psi^j\varepsilon_{T+h-i+j}\leq  ub_{T+h}\bigg)\\
        &-\mathbf{1}\bigg(\pi_T+\sum_{i=0}^h\sum_{j=0}^{M-h-i}\phi^i\psi^j\varepsilon_{T+h-i+j}\leq  lb_{T+h}\bigg)\Bigg] 
    \end{aligned}
    $}
    \label{bra-eq:sim.estimator_2}
\end{equation}

Given the information set known at time \textit{T}, the indicator functions in \eqref{bra-eq:sim.estimator_2} are only functions of the \textit{M} future errors, $\varepsilon_+^*=\Big(\varepsilon_{T+1}^{*},\dots,\varepsilon_{T+M}^{*}\Big)$. Let $q(\varepsilon_+^*)$ be the function providing the value of the difference between the two indicator functions. Furthermore, let $\varepsilon_{+}^{*(j)}=\Big(\varepsilon_{T+1}^{*(j)},\dots,\varepsilon_{T+M}^{*(j)}\Big)$, with $1\leq j \leq N$, be the \textit{j}-th simulated series of \textit{M} independent errors, randomly drawn from the errors distribution, here a Student's \textit{t}(3.25). Assuming that the number of simulations \textit{N} and the truncation parameter \textit{M} are large enough, the probability that the inflation rate (which follows an \textit{MAR}(1,1) process) will remain within the bounds in \textit{h} months can be approximated as such \citep{lanne2012optimal},\footnote{See Appendix \ref{bra-app:lanne} for more detailed derivations of the estimator.} 

\begin{equation}
\resizebox{0.93\textwidth}{!}{$
    \begin{aligned}
        \mathbb{P}\Big(lb_{T+h}\leq\pi_{T+h}^*\leq ub_{T+h}\big|\mathcal{F}_T\Big)&\approx\mathbb{E}_T\bigg[q\big(\varepsilon^*_+\big)\bigg]\\
        &\approx \frac{N^{-1}\sum_{j=1}^N q\Big(\varepsilon_{+}^{*(j)}\Big)g\Big(u_T-\sum_{i=1}^M\psi^i\varepsilon^{*(j)}_{T+i}\Big)}{N^{-1}\sum_{j=1}^Ng\Big(u_T-\sum_{i=1}^{M}\psi^i\varepsilon^{*(j)}_{T+i}\Big)},
    \end{aligned}
    $}
    \label{bra-eq:sim.estimator_3}
\end{equation} 
where $g$ is the \textit{pdf} of the Student's \textit{t}(3.25) distribution. \\

\citet{voisin2019forecasting} results show that with Cauchy-distributed errors, this approach is a good estimator of theoretical probabilities but are significantly sensitive to the number of simulations \textit{N} during locally explosive episodes. For Student's \textit{t} distributions however, results cannot be compared to theoretical ones, but as the number of simulations gets larger, the derived densities converge to a unique function. Overall \citet{voisin2019forecasting} show that: \textit{(i)} for degrees of freedom close to 3 and \textit{(ii)} during stable episodes, this approach yields consistent results that are not significantly sensitive to the number of simulations, as long as it is reasonably large. Hence, choosing the right number of simulations per iteration should not be a worry in this analysis.

\subsection{Sample-based approach}
As an alternative to using simulations, \citet{filtering} employ all past observed values of the process to approximate the marginal distributions of noncausal processes. They propose the following sample-based approximation of the predictive density of an MAR(0,1),\footnote{See Appendix \ref{bra-app:GJ} for more detailed derivations of the estimator.}

\begin{equation}
    \begin{split}
        &{l}(u^*_{T+1},\dots,u^*_{T+h}|\mathcal{F}_{T})\\ 
        & \approx {g}({u}_{T}-\psi u^*_{T+1})\dots {g}(u^*_{T+h-1}-\psi u^*_{T+h})
        \frac{\sum_{i=2}^{T}
        {g}(u^*_{T+h}-{\psi} u_i)}{\sum_{i=2}^{T} 
        {g}(u_{T}-{\psi} u_i)},
    \end{split}
    \label{bra-eq:samplebasedestim_u}
\end{equation}
where \textit{g} is the \textit{pdf} of a Student's \textit{t} distribution with 3.25 degrees of freedom. \\

Given a correctly identified model and based on the equivalence of the information sets $(\pi_1,\dots,\pi_T, \pi^*_{T+1},\dots,\pi^*_{T+h})$ and $(v_1,\varepsilon_{2},\dots,\varepsilon_{T-1},u_{T},u^*_{T+1},\dots,u^*_{T+h})$ \citep{filtering}, where $v_t=\pi_t-\psi \pi_{t+1}$, the predictive density of the \textit{MAR}(1,1) process $\pi_t$ can be obtained by substituting the filtered noncausal process $u_t$ by the mixed process $\pi_t$ in \eqref{bra-eq:samplebasedestim_u}, \\
\begin{equation}
    \begin{split}
    {l}(\pi^*_{T+1},\dots,&\pi^*_{T+h}|\mathcal{F}_T)\approx {g}\big((\pi_T-\phi \pi_{T-1})-\psi(\pi^*_{T+1}-\phi \pi_T)\big)\times \dots \\
    & \dots \times {g}\big((\pi^*_{T+h-1}-\phi \pi^*_{T+h-2})-\psi (\pi^*_{T+h}-\phi \pi^*_{T+h-1})\big)\\
    & \times \frac{\sum_{i=2}^{T}{g}\big(\pi^*_{T+h}-\phi \pi^*_{T+h-1}-{\psi} (\pi_i-\phi \pi_{i-1})\big)}{\sum_{i=2}^{T} {g}\big(\pi_T-\phi \pi_{T-1}-{\psi} (\pi_i-\phi \pi_{i-1})\big)}.
    \end{split}
    \label{bra-eq:samplebasedestim}
\end{equation}

Evidently, evaluating the conditional joint density over all possible outcomes becomes considerably computationally demanding as the forecast horizon increases. This is why \citet{filtering} developed a Sampling Importance Resampling (henceforth SIR) algorithm to counter this computational limitation. We provide details and describe the algorithm in the subsequent Section. We will therefore employ estimator \eqref{bra-eq:samplebasedestim} for a forecast horizon of 1 (see \eqref{bra-eq:samplebasedestim_h1}) and will use the SIR algorithm for horizons of 3 and 6 months.

\begin{equation}
\resizebox{0.91\textwidth}{!}{$
        l\big(\pi^*_{T+1}|\mathcal{F}_T\big)
         \approx g\big((\pi_T-\phi \pi_{T-1})-\psi(\pi^*_{T+1}-\phi \pi_T)\big)
        \frac{\sum_{i=2}^T g\big(\pi^*_{T+1}-\phi \pi^*_T-\psi (\pi_i-\phi \pi_{i-1})\big)}{\sum_{i=2}^T g\big(\pi_T-\phi \pi_{T-1}-\psi (\pi_i-\phi \pi_{i-1})\big)}.
        $}
    \label{bra-eq:samplebasedestim_h1}
\end{equation}

This estimator provides predicted probabilities that are a combination of theoretical probabilities and probabilities induced by past events; results are therefore case-specific and are based on a learning mechanism \citep{voisin2019forecasting}. For values close to the median this approach provides accurate and similar results to theoretical probabilities (when available) and to the simulations-based method of \citet{lanne2012optimal} for one-step ahead forecasts. Discrepancies widen as the level of the series increases. \\

\subsection{Sampling Importance Resampling algorithm}
As previously mentioned, estimator \eqref{bra-eq:samplebasedestim} developed by \citet{filtering} is substantially computationally demanding to employ for long forecast horizons and simulating from this distribution is rather intricate. The authors then proposed a SIR algorithm to counter this. The algorithm consists in simulating potential paths of future noncausal components $u_t$'s from an instrumental misspecified model from which it is easier to simulate. The distribution \eqref{bra-eq:samplebasedestim} of interest is then recovered using a weighted resampling of the simulations and the relation \eqref{bra-eq:MAR.filter_u} between the inflation rate $\pi_t$ and its noncausal component $u_t$. \citet{gourieroux2013explosive} note that a Markov process in reverse time is also a Markov process -- of the same order -- in calendar time with non-linear dynamics. Hence, since in this analysis $u_t$ is a non-causal MAR(0,1) process, it could be expressed as a causal AR(1) process, with non-linear dynamics. For a study on misspecified causal analysis of noncausal processes see \citet{gourieroux2018misspecification}. Following \citet{filtering}, we employ a Gaussian AR(1) model as instrumental model for the algorithm, 
\begin{equation}
    {u}_t=\tilde{\rho}{u}_{t-1}+\tilde{\varepsilon_t}.
    \label{bra-SIR}
\end{equation}
The parameter $\tilde{\rho}$ is estimated using standard OLS on the observed values ${u}_t$ filtered from the initial $MAR(1,1)$ process $\pi_t$. The errors $\tilde{\varepsilon}_t\sim IIN(0,\hat{\sigma}^2)$, where $\hat{\sigma}^2$ is the MAR residuals variance, $\tilde{\cdot}$ indicates estimation from the instrumental model, and $\hat{\cdot}$ from the initial $MAR(1,1)$ model. The conditional predictive density for the instrumental process is as follows, 
\begin{equation}
    \begin{split}
    & \tilde{F}({u}^*_{T+1},\hdots,{u}^*_{T+H}|{u}_{T}) \\
    & = \tilde{l}({u}^*_{T+H}|{u}^*_{T+H-1})\tilde{l}({u}^*_{T+H-1}|{u}^*_{T+H-2})\hdots \tilde{l}({u}^*_{T+1}|{u}_{T}) \\
    &=f({u}^*_{T+H}-\tilde{\rho}{u}^*_{T+H-1})f({u}^*_{T+H-1}-\tilde{\rho}{u}^*_{T+H-2})\hdots f({u}^*_{T+1}-\tilde{\rho}{u}_{T}),
    \end{split}
    \label{bra-eq:SIR_distrib}
\end{equation}
where $\tilde{F}$ is the predictive conditional distribution of \textit{h} future $u_t$'s from the instrumental model and $f$ the \textit{pdf} of a normal distribution with mean zero and variance $\hat{\sigma}^2$. Even though this model is clearly misspecified, the resampling step should automatically correct for the induced misspecifications \citep{filtering}. The algorithm for $h$-step ahead predictions is as follows \citep[see also][]{gourieroux2021forecast},
\begin{enumerate}
    \item \textbf{Sampling:} Draw $K$ series of $h$ independent values $\tilde{\varepsilon}$ from a normal distribution $N(0,\hat{\sigma}^2)$. Using the recursive equation \eqref{bra-SIR} and the last observed value $u_T$, compute the $K$ simulated paths $\big(\tilde{u}^i_{T+1},\hdots,\tilde{u}^i_{T+H}\big)$, respectively stacked in $\tilde{U}^i$'s, $i=1,\hdots,K$. 
    \item \textbf{Importance:} Denote $\Pi(\tilde{U}^i)$ the conditional density \eqref{bra-eq:samplebasedestim} evaluated at the path $\big(\tilde{u}^i_{T+1},\hdots,\tilde{u}^i_{T+H}\big)$. Compute the weights $w_i=\hat{\Pi}(\tilde{U}^i)/\tilde{F}(\tilde{U}^i)$ with $i=1,\hdots,K$. Namely, the ratio between the value of the density the algorithm intends to recover and the value of the instrumental density.
    \item \textbf{Resampling:} Draw with probability weighting based on the previously computed weights and replacement \textit{S} paths $\big(\tilde{u}^s_{T+1},\hdots,\tilde{u}^s_{T+H}\big)$, $s=1,\hdots,S$, from the $K$ simulated series in the sampling step, with respective weights $w_i$.\\ 
\end{enumerate}
Once the set of $S$ re-sampled $\tilde{U}^i$'s is obtained, each simulated path $\big(\tilde{u}^s_{T+1},\hdots$ $,\tilde{u}^s_{T+H}\big)$, $s=1,\hdots,S$ can be transformed into the corresponding future path for the variable of interest $(\pi^*_{T+1},\hdots,\pi^*_{T+H})$ using on the causal relation in Equation \eqref{bra-eq:MAR.filter_u}.\\

As stated by \citet{filtering}, if the number of initial simulations $K$ is large enough, simulating from the instrumental Gaussian model and applying the SIR algorithm is equivalent to simulating directly from the distribution of interest. It avoids simulating from a too complicated distribution, or avoids the simulation of too many extreme values. \citet{voisin2019forecasting} find that this approach may not function well during extreme events as the instrumental density becomes too different from the one to recover. The instrumental density (derived from normally distributed errors) is unimodal while the predictive density of a purely noncausal process during an explosive episode is bi-modal with a significantly larger range. This implies that during extreme events the instrumental conditional density is sometimes zero where the target distribution is not. This implies that during extreme episode the distribution derived from the algorithm may not converge to the target one since parts of the distribution are not simulated in the first step of the algorithm, regardless of the number of simulations. Further research should be done to improve the algorithm in such cases. However, we focus here on stable periods and this limit of the algorithm does therefore not affect our analysis.


\section{Forecasting with MAR model} \label{bra-sec:Application}
Using the model estimated in Section \ref{bra-sec:MAR}, we perform pseudo-real-time 1-, 3- and 6-months forecasts of the probabilities to remain within the bounds with expanding window from November 2016 to January 2020.  We hence put emphasis on short-term predictions which could provide warnings that the year-on-year inflation will be outside the target bounds in the near future. We first analyse the impact of the choice of forecasting method as well as the impact of the forecast horizon on the probabilities. As shown previously in Figure \ref{bra-fig:exp_window_estimation}, the estimation of the model is stable and particularly so after 2016. Therefore, re-estimating the model with expanding window at each point of forecasts does not have a significant impact on the results and represents a good real-time forecasting analysis. We can hence obtain 39 pseudo real-time probability forecasts for each of the three horizons and each of the two methods. \\

\begin{figure}[h!]
\begin{subfigure}{0.33333\textwidth}
  \centering
  \includegraphics[width=\linewidth]{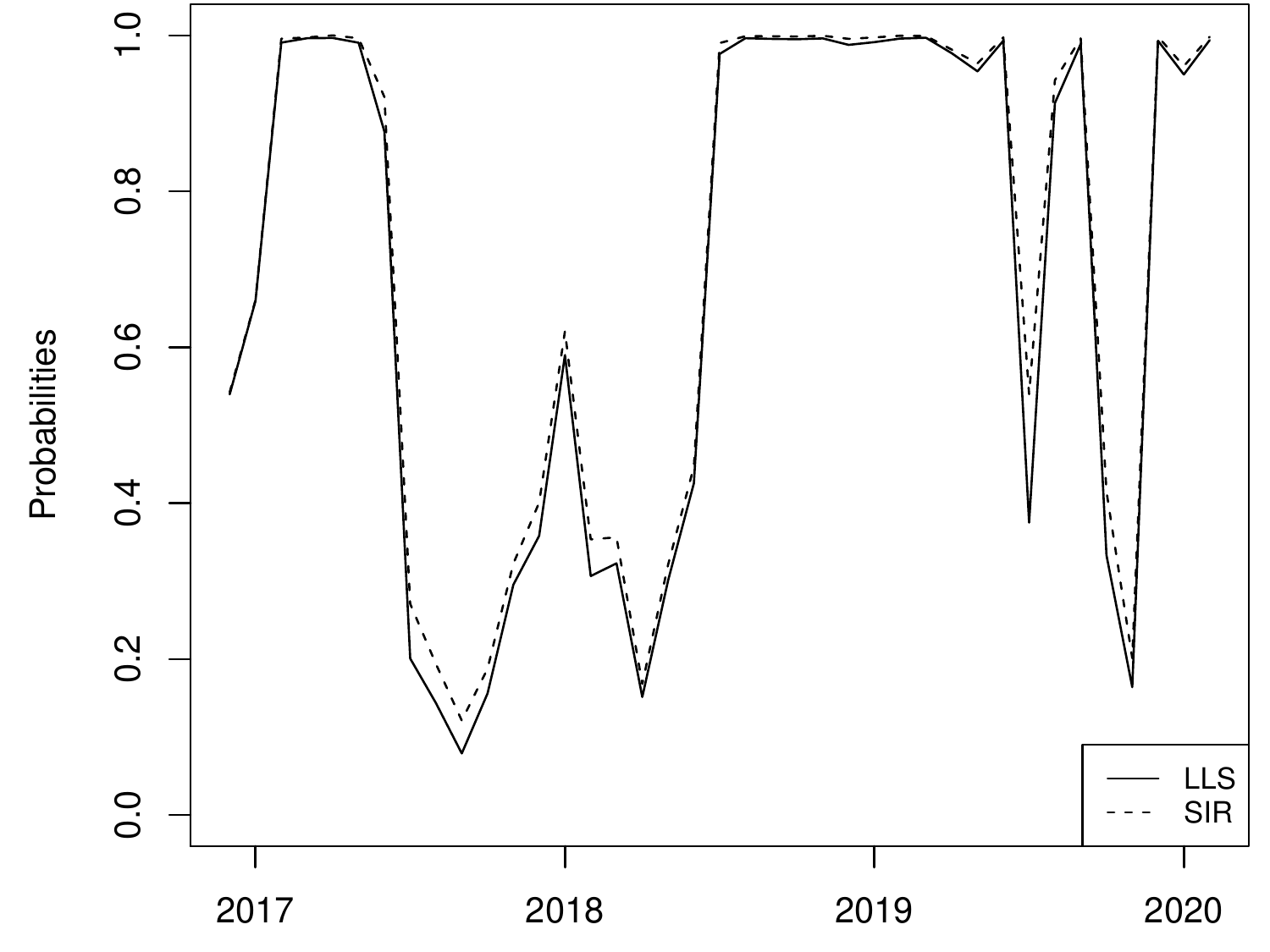}
  \caption{1-month ahead}
\end{subfigure}%
\begin{subfigure}{0.33333\textwidth}
  \centering
  \includegraphics[width=\linewidth]{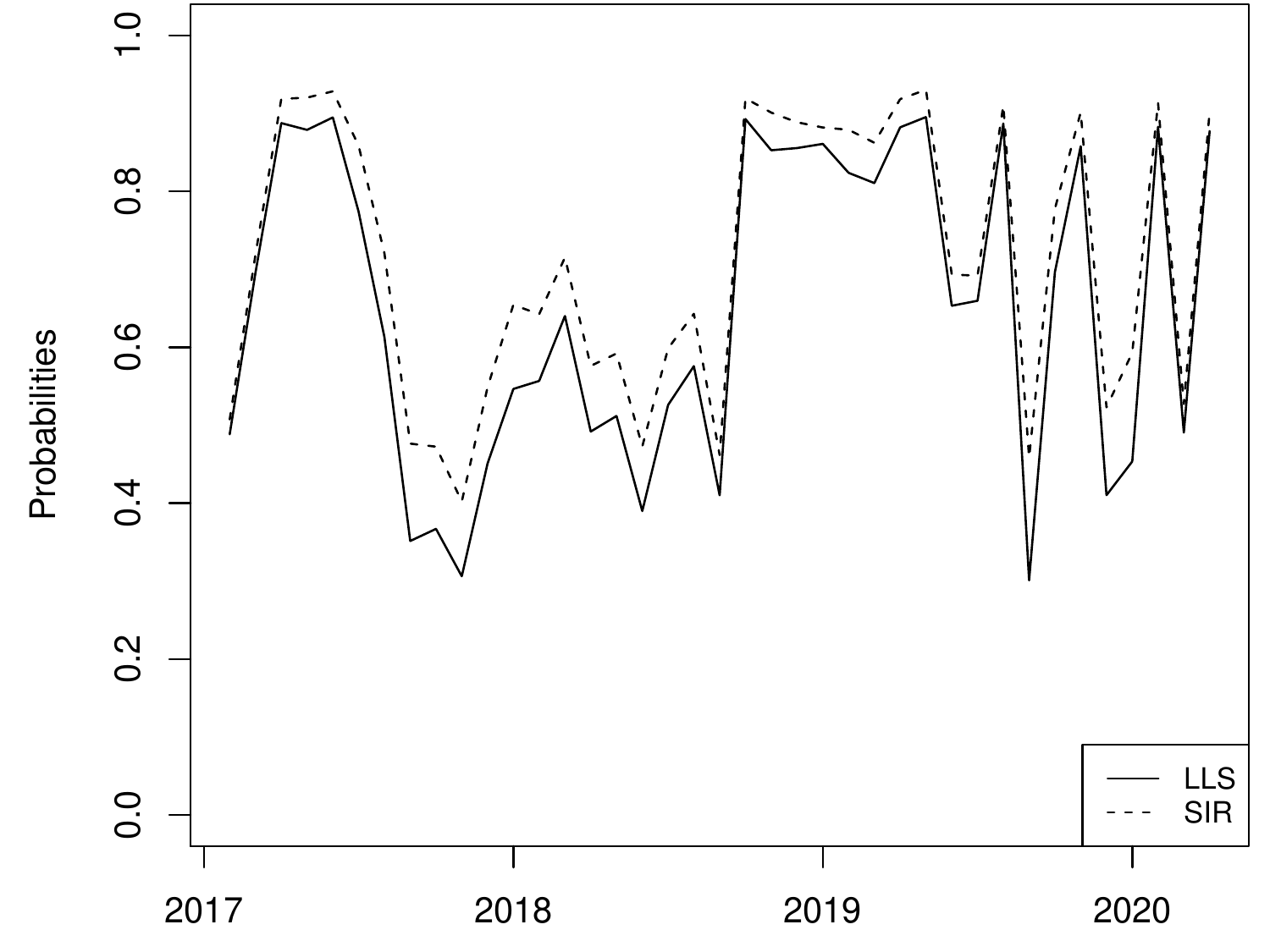}
  \caption{3-months ahead}
\end{subfigure}%
\begin{subfigure}{0.33333\textwidth}
  \centering
  \includegraphics[width=\linewidth]{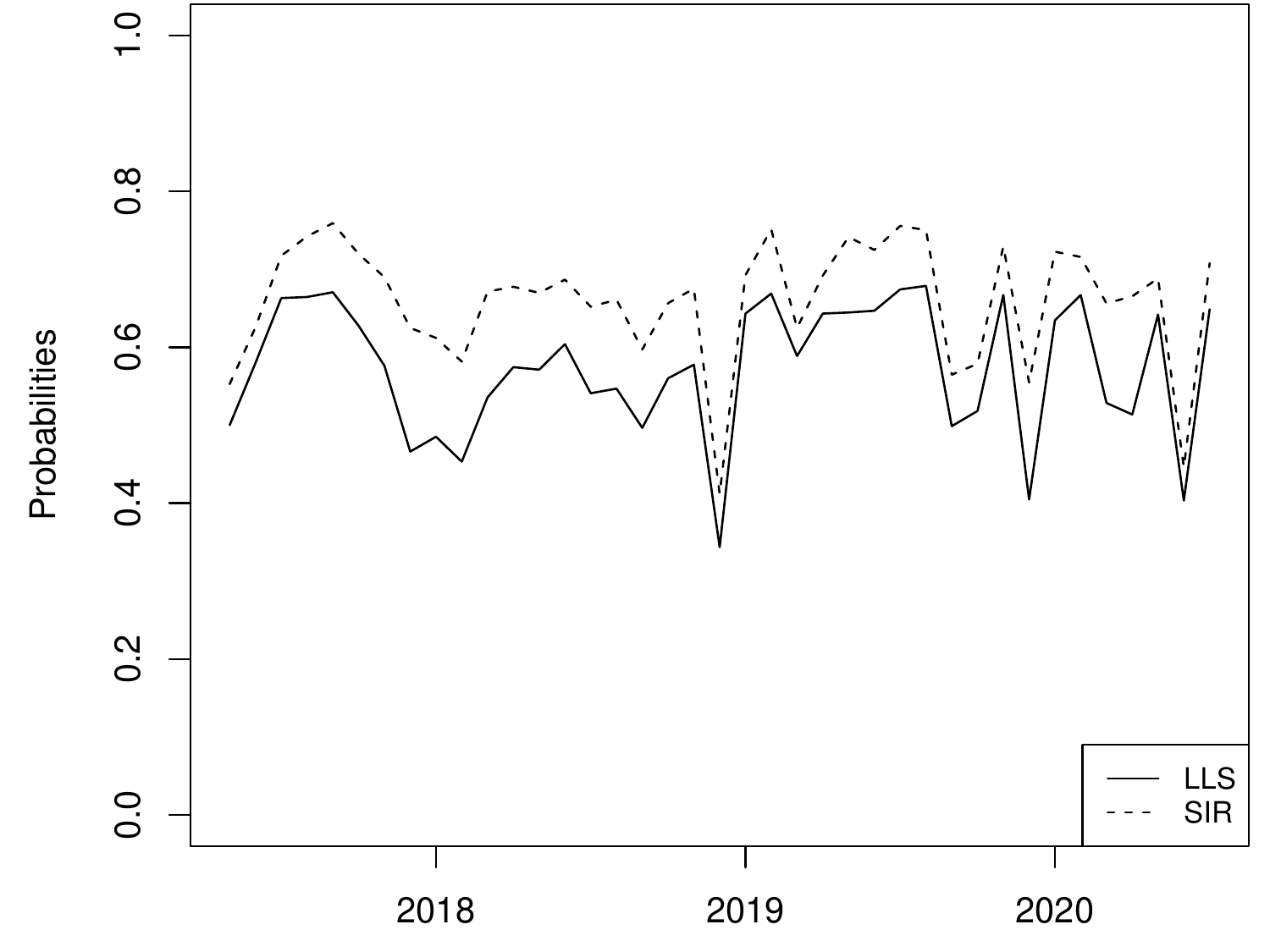}
  \caption{6-months ahead}
\end{subfigure}
\caption{Probability forecasts for inflation to remain within the bounds}
\label{bra-fig:predictions_both}
\end{figure}

Figure \ref{bra-fig:predictions_both} depicts the differences between the predictions made with the simulations based method (solid line) of \citet{lanne2012optimal}\footnote{Employing 1\,000\,000 simulations at each iteration.}, and the SIR algorithm (dashed line) employing the sample-based method of \citet{filtering}\footnote{Employing 100\,000 simulations in the first step and 10\,000 resampling forecasts.}. Graph (a) shows the 1-month ahead forecasts performed with increasing sample at each point. Graph (b) and (c) correspond to 3- and 6-months ahead forecasts respectively. All forecasts are performed at the same 39 points, starting in November 2016. Note that as the analysis does not regard extreme episodes, the 1-step ahead density forecasts obtained with the SIR algorithm and the sample-based method of \citet{filtering} were identical\footnote{Results available upon request.}. For a thorough analysis of their respective performance see \cite{voisin2019forecasting}. Since they perform similarly and because the SIR algorithm is less computationally demanding we only present the SIR results alongside the simulations-based results (LLS). \\

We can see that the differences between the two methods increase with the horizon. For one-step ahead forecasts, both approaches yield almost the same results at each point in time, while for the 6-months ahead forecasts they differ on average by 0.09, with the SIR probabilities always larger than LLS. This indicates that LLS predicts more volatile fluctuations in the inflation rate than the SIR for longer horizons, hence suggesting lower probabilities to remain within the bounds. Furthermore, we can see that while probabilities vary significantly at a 1-month horizon (from 0.08 to 1), they converge around 0.6 for both methods as the horizon increases to 6 months. That is, as the forecast horizon increases, the uncertainty prevails, regardless of being within or outside the bounds at the moment of the forecast, which yields almost 50/50 chance to meet the target at at 6-months horizon.\footnote{Table \ref{bra-tab:probas_in_bounds} in Appendix \ref{bra-sec:app.results} summarizes the results.} This is because we are investigating a period during which there is no explosive episodes. Inflation is relatively stable though close to the lower bound of the target. \\

\section{MARX model} \label{bra-sec:MARX}

\subsection{Model representation}

MAR models with additional strictly exogenous variables have been introduced by \citet{hecq2020mixed}. The so called MARX model allows to estimate the effects of covariates without aggregating them in the error term as in the MAR($r,s)$ model of Section \ref{bra-sec:MAR}. The MARX($r,s,q$) for a stationary time series $\pi _{t}$ (here, the annual IPCA inflation rate) reads as follows%
\begin{equation}
\phi (L)\varphi (L^{-1})\pi _{t}-\beta ^{\prime }X_{t}=\varepsilon _{t},
\label{bra-eq:MARX}
\end{equation}%
where $\phi (L)$ and $\varphi (L^{-1})$ are the lag and lead polynomials of order $r$ and $s$ with $r+s=p$ and $q$ is the number of strictly exogenous variables. We still assume that the roots of both polynomials lie outside the unit circle. When $q=0$, the process reduces to a standard MAR($r,s$). In case $q>0$, the process no longer has a strictly stationary solution solely in terms of $\varepsilon _{t}$, but involves additionally $X_{t}$ (consisting of $q$ exogenous variables). That is, 
\begin{equation}\label{bra-eq:marx.inverse}
\pi _{t}=\gamma (L,L^{-1})\varepsilon _{t}+\gamma (L,L^{-1})\beta ^{\prime
}X_{t}=\sum_{j=-\infty }^{\infty }\gamma _{j}z_{t-j}, 
\end{equation}%
where $z_{t-j}=\varepsilon _{t-j}+\sum_{i=1}^{q}\beta _{i}x_{i,t-j}$ and $\gamma (L,L^{-1})$ is an operator satisfying $\gamma (L,L^{-1})\phi(L)\varphi (L^{-1})=1$ such that $\pi _{t}$ has a two-sided MA-representation augmented with past, current and future values of $X_{t}$. Consequently, denoting $\tilde{x}_{i,t}=\beta _{i}x_{i,t}$, $\pi _{t}$ consists of a two-sided MA representation and the sum of $q$ processes $\tilde{x}_{i,t}$ that are passed through a two-sided linear filter with coefficients resulting from inverting the product $[\phi (L)\varphi (L^{-1})]$. Similarly to the MAR, the MARX can be decomposed in $u$ and $v$ components, namely the noncausal and causal components respectively, 
\begin{equation}\label{bra-eq:marx.u.inverse}
u_{t}\equiv \phi (L)\pi _{t}\leftrightarrow \varphi (L^{-1})u_{t}-\beta
^{\prime }X_{t}=\varepsilon _{t}, 
\end{equation}%
\begin{equation}
v_{t}\equiv \varphi (L^{-1})\pi _{t}\leftrightarrow \phi (L)v_{t}-\beta
^{\prime }X_{t}=\varepsilon _{t}. 
\end{equation}%
This $(u,v)$ representation is useful to simulate variable and to forecast MAR processes. See \citet{hecq2020mixed} for details on the model.

\subsection{Data}

We evaluate to what extent key drivers of inflation identified by the BCB, have an influence on the probability that inflation stays within the target bounds. The variables we consider are the year-on-year percentage change in industrial production\footnote{ We prefer to use the industrial production index as a measure of economic activity over the GDP that is available quarterly.} ($ip_{t})$, the year-on-year percentage change in the Real/US\$ exchange rate ($ex_{t})$ where the rate refers to the last working day of the period and the year-on-year percentage change in the Selic target interest rate ($ir_{t}$). We denote $X_{t}=(ip_{t},ex_{t},ir_{t})$ the stacked stationary exogenous variables at time \textit{t}. We use published indicators of the three variables (retrieved from FRED and from the Central Bank of Brazil databases). For coherence and comparison purpose we use the same sample as for Section \ref{bra-sec:MAR}, namely from January 1997 to January 2020.

\subsection{Estimation results}
The aim of this section is to analyse the added value of augmenting the MAR(1,1) model identified in Section \ref{bra-sec:Application} with exogenous variables. Hence, we identify the MARX(1,1,\textit{q}) model best fitting the data, using the strategy proposed in \citet{hecq2020mixed}. We first estimate an ARDL model using a maximum likelihood approach and find that BIC favors an ARDL(2,0,0,0), namely an AR(2) with only contemporaneous values of the three regressors. The increase of the goodness of fit is however not substantial, the $\bar{R}^{2}$ increases from 0.977 to 0.980. The estimated model is the following (HCSE standard errors in brackets),
\begin{equation*}\label{bra-eq:ARDL}
        {\pi}_{t}=\underset{\textit{(0.05)}}{1.46}\pi _{t-1}-\underset{\textit{(0.05)}}{0.50}\pi _{t-2}+\underset{\textit{(0.56)}}{2.21}ip_{t}+\underset{\textit{(0.17)}}{0.87}ex_{t}+\underset{\textit{(0.09)}}{0.23}ir_{t}+\eta_t.
\end{equation*}

All three regressors have a significant impact on annual
inflation and only their contemporaneous values are selected. The value of the Jarque and Bera normality test is 185.3 with a $p-value <$ 0.0001. \\

Following the methodology described in \citet{hecq2020mixed} we compare all MARX(1,1,3) considering all possible time indices of the exogenous variables (namely their lag, lead or contemporaneous value). The model maximizing the likelihood function is the following MARX(1,1,3) with a $t(3.39)$ (standard errors in brackets),

\begin{equation}\label{bra-eq:marx_estimation}
    \resizebox{0.915\textwidth}{!}{
        $(1-\underset{\textit{(0.04)}}{0.50}L)(1-\underset{\textit{(0.01)}}{0.97}L^{-1}){\pi}_{t}=\underset{\textit{(0.38)}}{1.64}ip_{t+1}-\underset{\textit{(0.09)}}{0.53}ex_{t+1}-\underset{\textit{(0.08)}}{0.04}ir_{t+1}+\varepsilon_t.$
    }
\end{equation}
That is, the model maximizing the likelihood function includes the lead of the three exogenous variables. The year-on-year inflation rate is therefore influenced by anticipations in the key economic variables considered here. The coefficients might not represent actual effects but instead corrections of the effect already captured by the coefficient on the lead of the inflation rate. We do not focus on that.\footnote{We acknowledge the fact that these regressors might not be fully strictly exogenous. Yet, we provide this analysis to illustrate the use of MARX models and take advantage of the availability of frequent forecasts for these variables which enables using MARX models to forecast without needing to predict the exogenous variables separately.} We can see that the lag coefficient has slightly decreased (by 0.04) while the lead coefficient is now slightly higher (by 0.03).  \\

\subsection{Predictions with Focus data}
\citet{hecq2020mixed} find that the forecasts obtained in a MARX are superior to the ones of a MAR model when the future values of the regressors are known. The gain diminishes when the regressors must be forecasted as well, using an ARMA model for instance. The choice of exogenous variables in this analysis was motivated by the fact that those variables are forecasted and updated on a daily basis by experts in the Focus database  maintained by the Brazilian Central Bank.\footnote{See https://www.bcb.gov.br/en/monetarypolicy/marketexpectations. The Brazilian survey on economic forecasters is unique. Everyday, a set of experts (from banks, fund managers, brokers, consulting companies, etc) give their evaluation on the future of inflation rate as well as for several key macroeconomic variables regarding the Brazilian economy. Most variables are published before the release of the Brazilian consumer price index, which is made public around the 10th of the subsequent month. Some variables have some delays and they also often continue to be forecasted for several vintages after the end of the corresponding month.} Previous periods are re-evaluated until the official numbers are published and forecasts for future periods are available at various horizons. This allows to perform forecasts using MARX models without modelling and forecasting separately the exogenous variables. \\

In this study, when performing predictions, we take the overall median of experts' forecasts for the future values of $X$ in real time. This means for instance that in May 2019 we take the forecasts of the explanatory variables for the next months made at the end of May 2019. Furthermore, to ensure coherence in the data, we also replace the last three data points (in this example we replace March, April and May 2019) by the last vintages available at the point at which the forecast is preformed. \\

Alike the MARX process $\pi_t$, the MA representation of the noncausal component $u_t$ is also augmented by contemporaneous and future values of $X_t$,
\begin{equation}\label{bra-eq:marx.inverted}
\begin{split}
    u_{t}&=\varphi (L^{-1})^{-1}\big[\beta^{\prime }X_{t}+\varepsilon _{t}\big]\\
    &=\sum_{i=0}^\infty \psi^i [\beta^{\prime }X_{t+i}+\varepsilon _{t+i}\big].
\end{split}
\end{equation}
In analogy to Equation \eqref{bra-eq:u.in.eps}, the method of \citet{lanne2012optimal} -- if possibly extended to MARX models -- would require a truncation parameter \textit{M} large enough to approximate \eqref{bra-eq:marx.inverted}. This implies that \textit{M} future values of the exogenous variables would be needed to perform forecasts with such approach. \citet{lanne2012optimal} suggest using \textit{M}=50 which represents more than four years of monthly predictions.\\

On the other hand, the method of \citet{filtering} only requires as many predictions of the exogenous variables as the forecast horizon. This makes this approach the most suitable for this analysis. As explained in Section \ref{bra-sec:Application}, for one-step ahead forecasts, the estimator \eqref{bra-eq:samplebasedestim_h1} can be employed, but as the forecast horizon increases the SIR algorithm alleviates the computational limitations of the estimator. The SIR approach simulates the noncausal component of the process using a purely causal instrumental model and then transforms those simulations into simulations of the variable of interest using the causal relation between $\pi_t$ and $u_t$ described in \eqref{bra-eq:MAR.filter}. \\

We use the following instrumental model for the SIR algorithm,
\begin{equation}\label{bra-eq:marx_pseudo_causal}
u_t= \rho u_{t-1} + \eta_1 ip_{t} +\eta_2 ex_{t} +\eta_3 ir_t + \epsilon_t,
\end{equation}
where $\epsilon_t\sim N(0,{\sigma}^2)$. This instrumental model is the pseudo causal model obtained by inverting the time indices in the noncausal representation of $u_t$. The parameters are obtained using standard OLS on the whole sample. \\

As mentioned, to resemble a real time forecast, we replace the values for the last three observations (from \textit{T}-2 to \textit{T}) by the evaluations made by experts for the exogenous variables at time \textit{T}, and we take the six consecutive months forecasts made at the same point in time. Alike Section \ref{bra-sec:Application}, we perform 1-, 3- and 6-months ahead forecasts at the end of each month from November 2016 to January 2020. The SIR algorithm is employed the same way as for the MAR. We re-estimate the model every time, yet we fix the time indices of the variables to be the ones in model \eqref{bra-eq:marx_estimation}.\footnote{Note that estimation is stable for all coefficients and predictions are thus easily comparable from one point in time to another.} \\

\begin{figure}[h!]
\begin{subfigure}{\textwidth}
  \centering
  \includegraphics[width=0.8\linewidth]{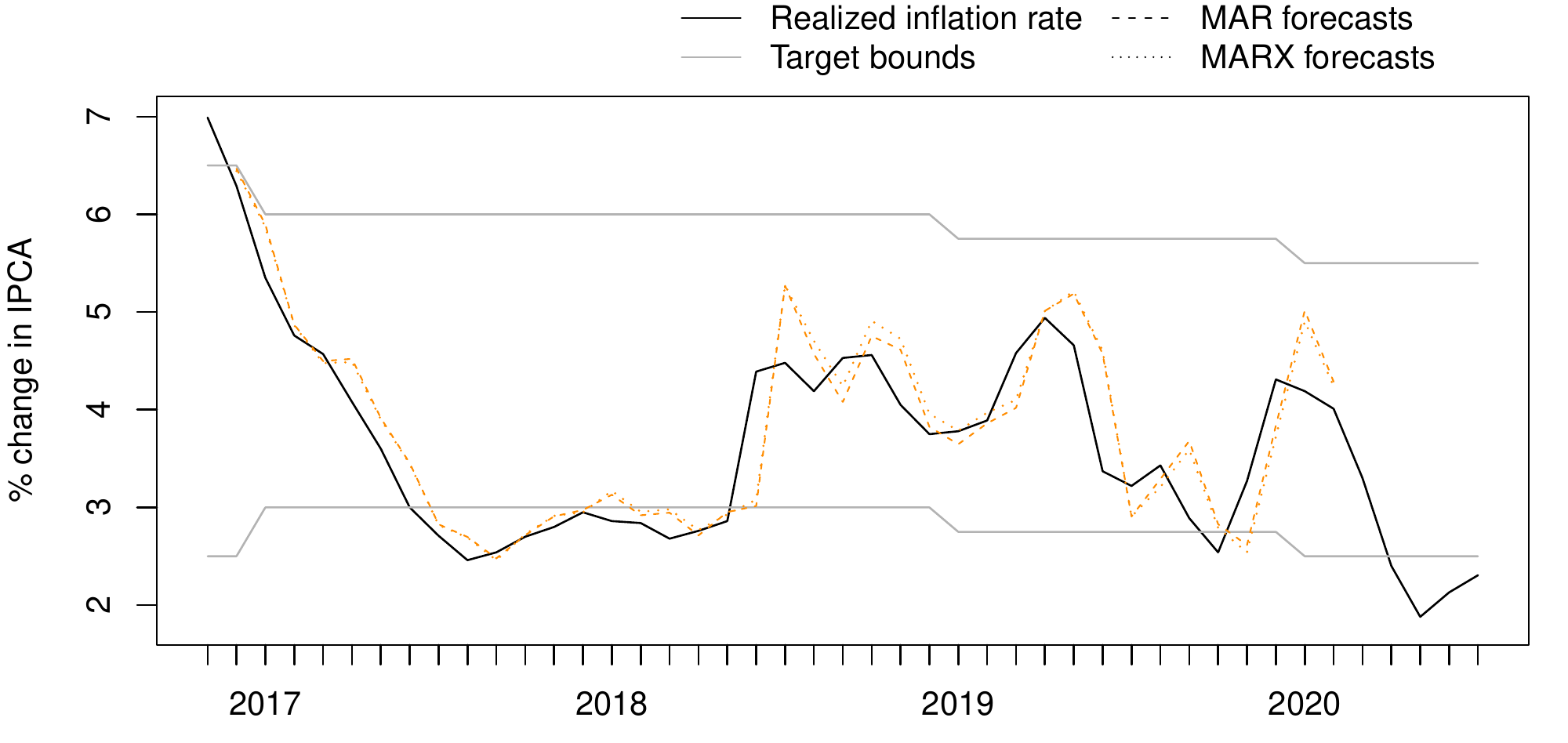}
  \caption{1-month ahead}

\end{subfigure}%
\newline
\begin{subfigure}{\textwidth}
  \centering
  \includegraphics[width=0.8\linewidth]{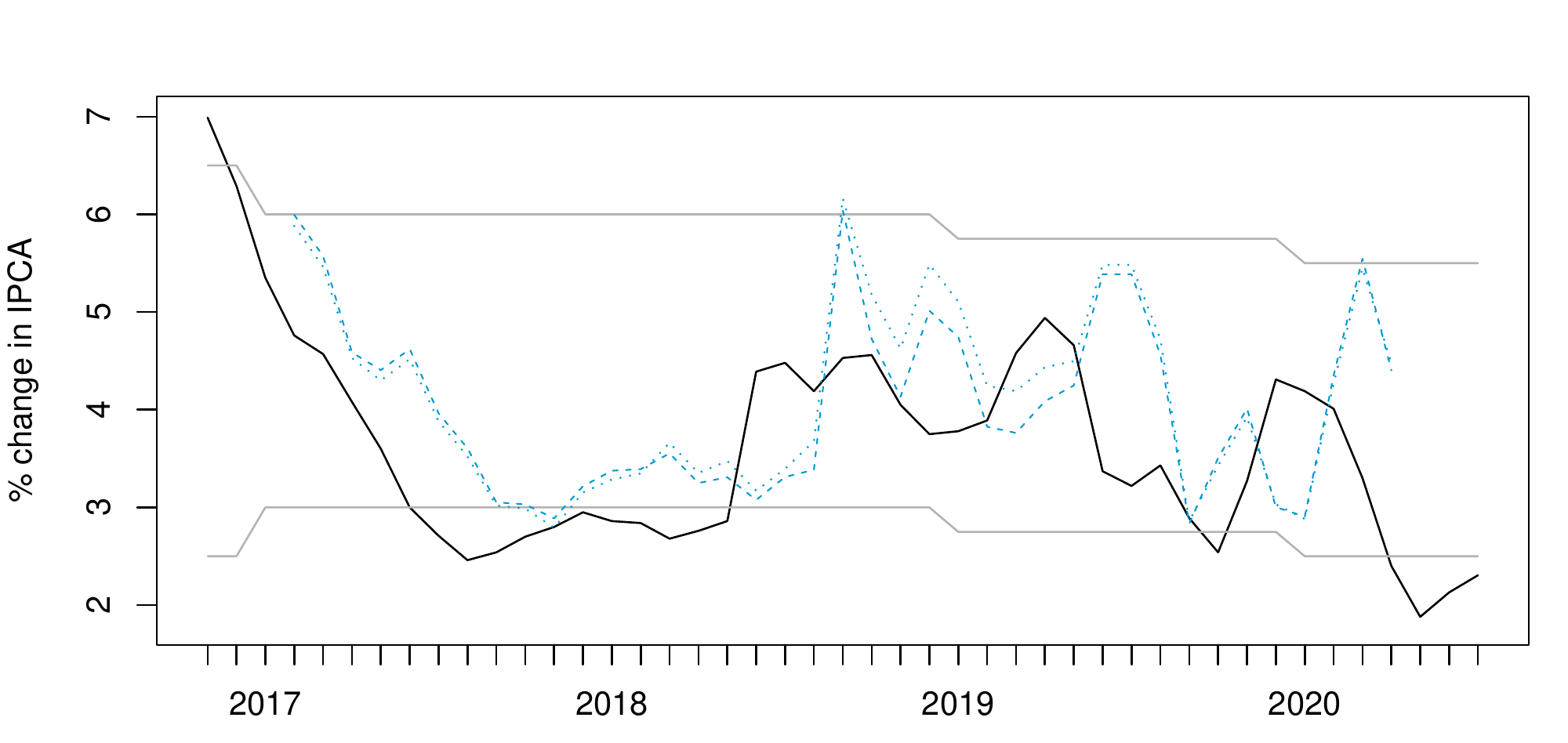}
  \caption{3-month ahead}

\end{subfigure}
\newline
\begin{subfigure}{\textwidth}
  \centering
  \includegraphics[width=0.8\linewidth]{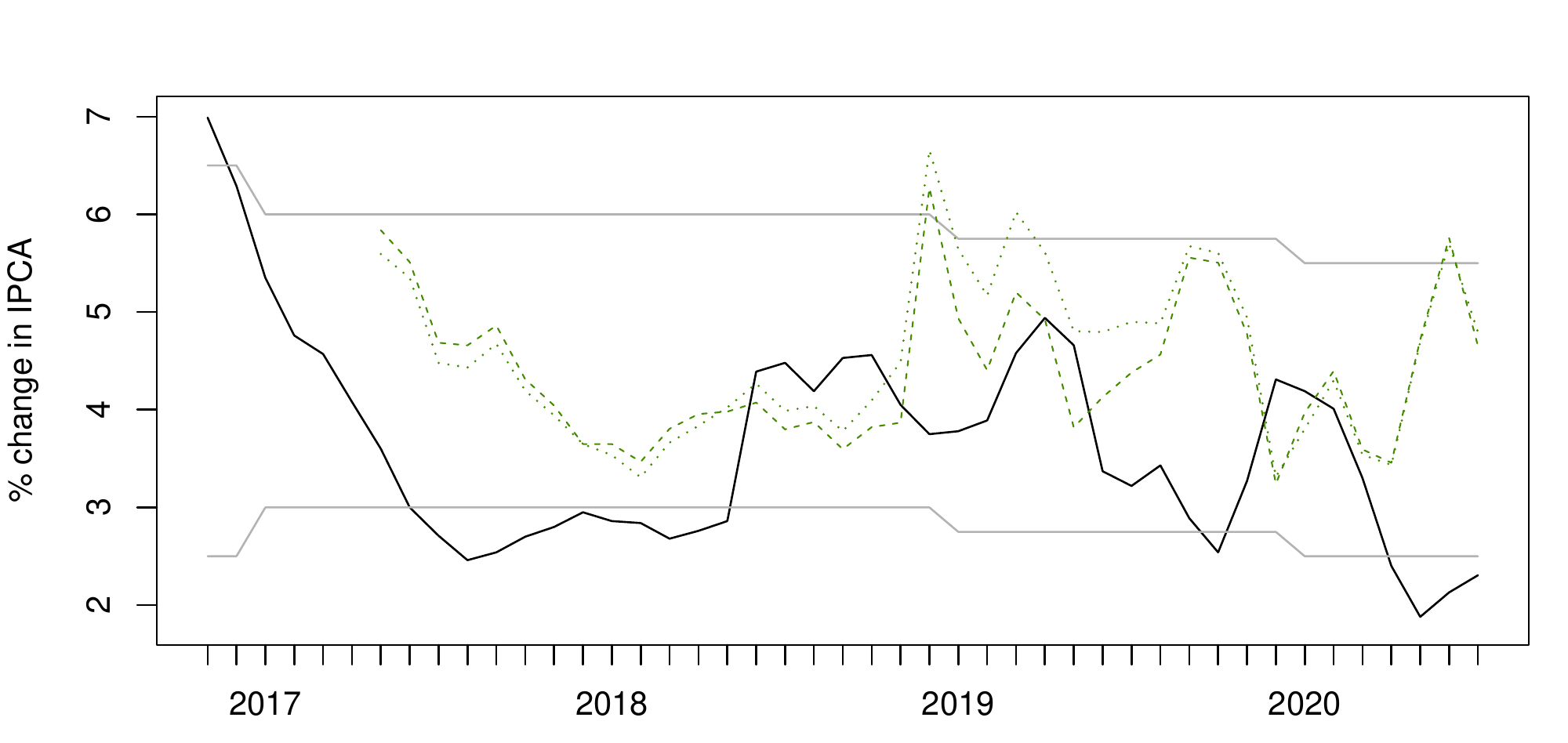}
  \caption{6-month ahead}

\end{subfigure}
\caption{MAR vs MARX}
\label{bra-fig:MAR_vs_MARX_pointforecasts}
\end{figure}

Figure \ref{bra-fig:MAR_vs_MARX_pointforecasts} shows point forecasts obtained from the MAR model of Section \ref{bra-sec:Application} (dashed lines) compared to the point forecasts obtained with the MARX model (dotted lines). Forecasts at the different horizons are compared to realized inflation rate (black solid line) and the announced target bounds (grey solid lines). We can see that the inclusion of exogenous variables does not significantly alter the predictions, especially for short-term horizons. An explanation to this is that forecasts of yearly changes in the exogenous variables around that time were very close to 0 as we are investigating a stable period. We can however notice that when the inclusion of exogenous variables has an impact, it becomes increasingly noticeable at larger horizons. We can therefore expect that during unstable and more volatile periods the inclusion of exogenous variables should influence more significantly forecasts and predictive densities. Another explanation is that inflation might already be influenced by expectations of the exogenous variables. Hence, their inclusion might most of the time not be enough to significantly alter predictions. 

\section{Short-term credibility}\label{bra-sec:Credibility}
There is a long tradition of evaluating the credibility of central banks. \citet{svensson2000should} for instance measures credibility as the distance between expected and targeted inflation. A commonly used definition of central bank credibility is the following: \textit{a central bank is credible if people believe it will do what it says} \citep{blinder2000central}. This definition however implies the need for modelling people’s beliefs, which is not straightforward and can induce significantly different results based on the assumptions and model construction (see the reference list in \citealp{issler2019central}). Some construct the credibility as an inverse function of the gap between expectation and target, using different ad-hoc thresholds for what is considered credible or not (see among others \citealp{cecchetti2002central} and \citealp{mendoncca2007credibilidade}). \citet{bomfim2000opportunistic} construct expected inflation as a weighted average of the target and past inflation rate and interpret credibility as the weight of the latter. \citet{dovern2012disagreement} use professional forecasters’ predictions and interpret the discrepancies between them as an indication of lack of credibility. \citet{issler2019central} propose a bias-corrected measure of inflation expectation using survey data and construct a credibility index based on whether the target falls within the confidence interval of their expectation measure. Most credibility measures are constructed for long-term horizons, such as 12 months ahead for instance, so that short-term shocks to inflation vanish. \\

We take a different stand-point and measure credibility not from the perspective of people' beliefs but from the dynamics in past inflation rates, building on the forward-looking characteristics of MAR models. That is, we use the probabilities that yearly inflation remains within the target bounds as an indication of whether the Central bank's target is currently credible or not. This measure therefore only relies on the statistical and dynamic properties of inflation and not on people’s beliefs. To retain and employ all the dynamics in realised inflation rates, we use monthly year-on-year inflation rate, which implies that our forecasts are based on shorter time horizons than the one-year-ahead expectations used in most of the aforementioned methods to build the credibility indices. We therefore measure short-term credibility and not long-term, which explains the discrepancies obtained with other approaches. Indeed, short-term predictions carry more shocks than one-year ahead forecasts. Our approach can be used in real time, here on a monthly basis, as an early warning that yearly inflation will exit the bounds in the near future. \\

\begin{figure}[h!]
    \centering
    \includegraphics[width=0.9\textwidth]{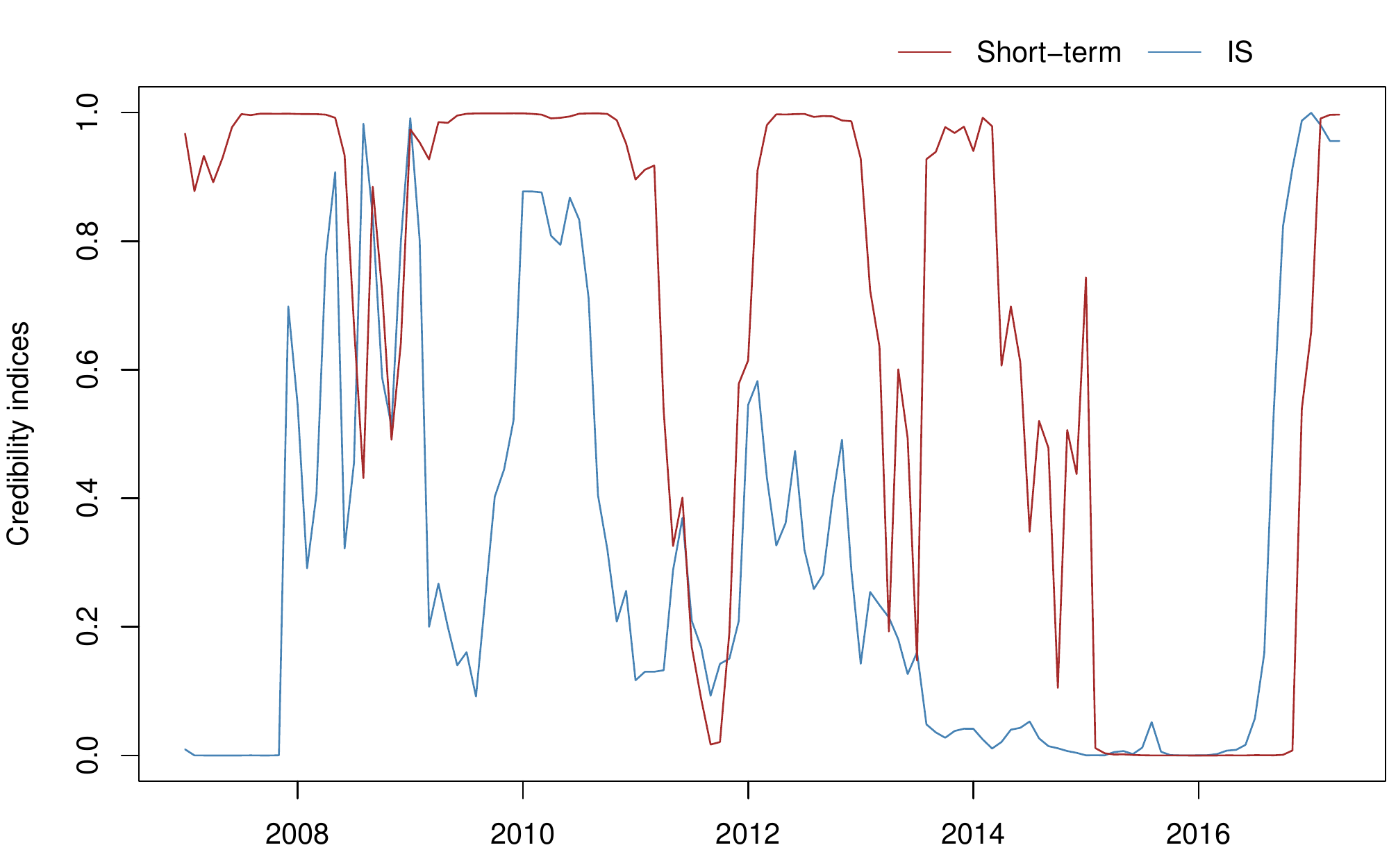}
    \caption{Comparison credibility indices}
    \label{bra-fig:CB_credibility}
\end{figure}

For comparison purposes we use the same time span as the analysis of \citet{issler2019central}, namely from January 2007 to April 2017. Figure \ref{bra-fig:CB_credibility} compares our short-term credibility measurement with the longer-term credibility index proposed in \citet{issler2019central} -- denoted as \textit{IS}. The short-term credibility index is the 1-month ahead simulations-based probabilities (LLS) that yearly inflation will remain within the target bounds (the red solid line), derived from the forward-looking MAR model. The discrepancies between the two methods stem from the horizon and perspective of each index. Recall that the long-term index is constructed a year ahead, that is, in 2013 for instance, people did not believe the Central bank would be credible in 2014. However, on a shorter-term basis, it seemed much more likely that the target would be met, hence a much larger short-term index. Our short-term approach does not investigate whether people trust the central bank to reach its goals in a year but instead whether it seems likely to happen at short horizons based on realized inflation rates. This makes the two approaches complementary. \\

\begin{figure}[H]
\begin{subfigure}{\textwidth}
  \centering
    \caption{Short-term index}
    \vspace{-0.5cm}
  \includegraphics[width=0.8\linewidth]{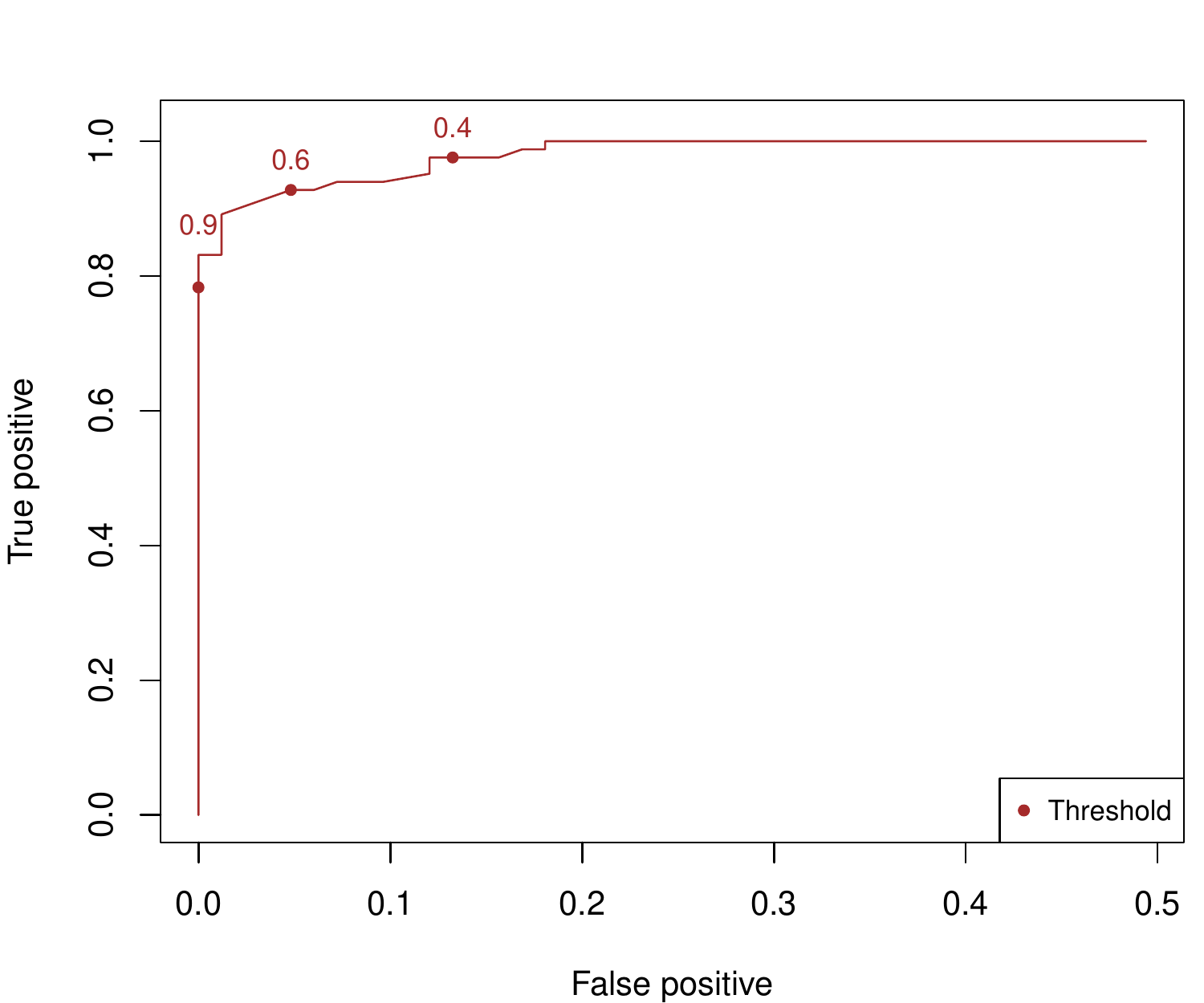}
\end{subfigure}%
\newline
\newline
\newline
\begin{subfigure}{\textwidth}
  \centering
    \caption{Long term indices}
  \includegraphics[width=0.8\linewidth]{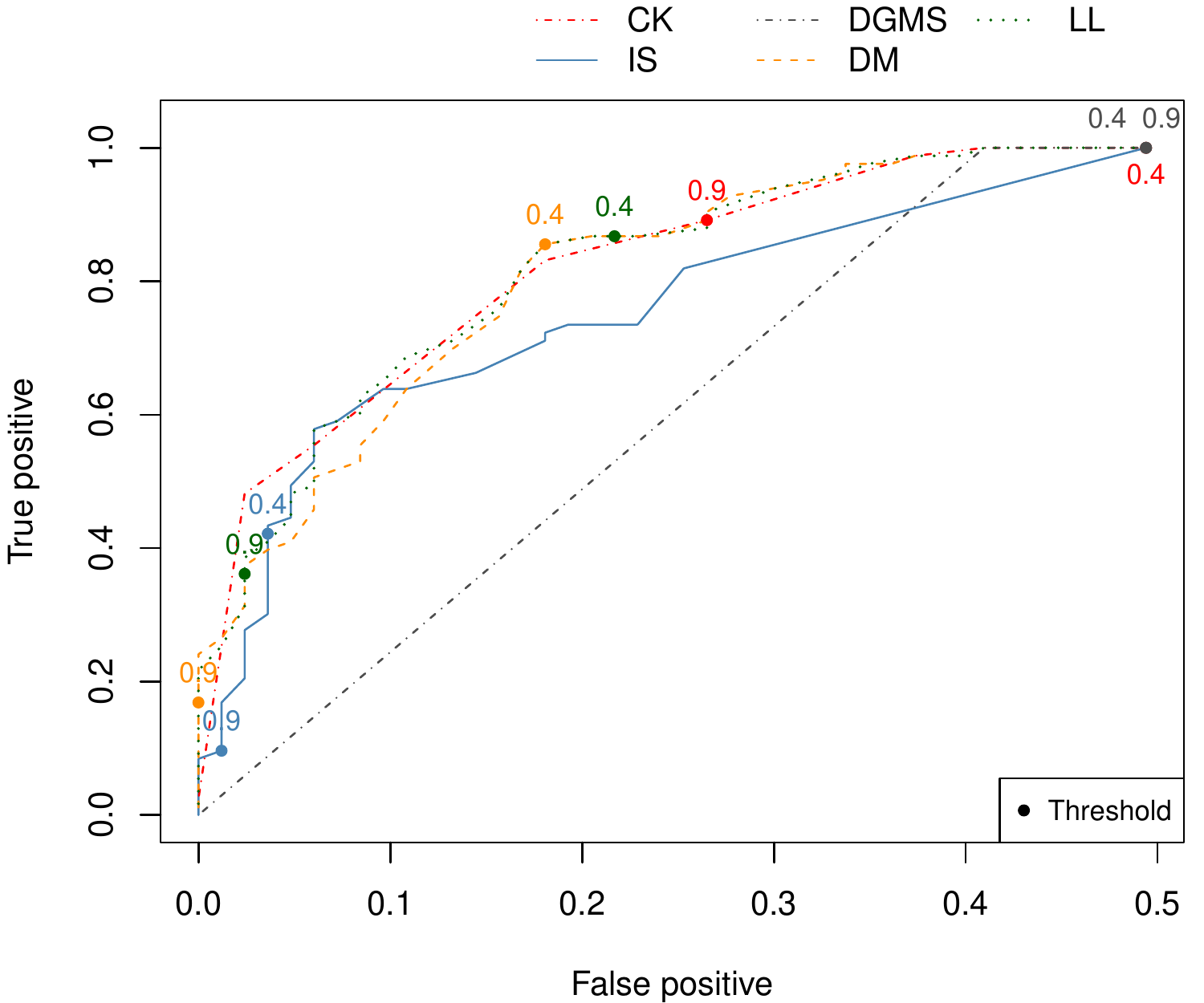}
\end{subfigure}
\caption{ROC curves}
    \label{bra-fig:ROC_curves}
\end{figure}

To provide a better understanding of the indices we construct the receiver operating characteristic (ROC) curve. It represents the accuracy of the credibility measures as the threshold from which we consider the Central bank to be credible varies. That is, for a given threshold \textit{x} and a given forecasted credibility index, the Central bank is predicted to be credible if the index is above \textit{x} and we compare it to the actual inflation rate in the same month. We can then compute the rates of true and false positives for each given threshold and index. A true positive is when the bank is predicted to be credible and the actual inflation rates turns out to be within the tolerance bounds and a false positive is when the bank was predicted to be credible but the inflation rate ended up outside the bounds. We consider the long term indices that part of the analysis by \citet{issler2019central}, namely the index they construct (IS), \textit{DM} is the index developed by \citet{mendoncca2007credibilidade}, \textit{CK} the index by \citet{cecchetti2002central}, \textit{DGMS} by \citet{de2009inflation} and \textit{LL} by \citet{levieuge2018central}. Figure \ref{bra-fig:ROC_curves} depicts the ROC curves for each of the indices. Graph (a) is our short-term index and graph (b) is for the long-term indices. It was expected to find that the short-term approach would outperform long-term forecasts in detecting when the inflation rate will exit the target bounds since predictions are made for a much closer horizon. For all indices, the ROC curves go from the top right corner when the threshold is set to 0 and to the bottom left corner when the threshold is set to 1. The sensitivity of each index to changes in the thresholds in between those extremes vary significantly from one to another. We see on graph (a) for the short-term index that thresholds between 0.4 and 0.9 provide a true positive rate above 0.8 and a false positive rate of less than 0.15. Hence this approach is not particularly sensitive to the choice of threshold for accurate results. The choice of threshold then only depends on the preference of the practitioner regarding the wanted true and false positive rates. We can however notice that long-term indices are much more sensitive to the choice of thresholds. We only depict the points for thresholds 0.4 and 0.9, but we can first notice how, for most methods, changes in thresholds in between those values will yield a very different results accuracy. Naturally, the longer horizon of these predictions imply a loss in precision, however, the results across long term indices are noticeably different. Some methods tend to overestimate the credibility, while others tend to underestimate it. If we compare our short-term index with the IS long-term index, which is not dependent on ad-hoc benchmarks or thresholds, we notice that people believe the Central bank to be much less credible that it actually is a yeah ahead. Indeed, any probability threshold in the ROC curve above 0.4 yield rather low true positive rates.  \\

Overall, while our short-term credibility index might be affected by short-term shocks in the inflation rate, it does not require any modelling of people's beliefs, or does not require any ad-hoc decisions of benchmarks. It can serve as an early warning of exiting the target bounds. The different perspective that our measure takes makes it complementary to longer-horizons credibility indices. Indeed, the long term credibility index built from people's beliefs combined with short-term probabilities to meet the target provide a clearer and more complete picture of the reliability of inflation targeting system.

\section{Conclusion}\label{bra-sec:Conclusion}
This paper investigates the probabilities that the Brazilian inflation rate remains within the target bounds determined by the central bank. We estimate mixed causal-noncausal models with leads and lags of the inflation rate and the first noticeable results are the stability of the parameters estimation over time. Contrary to usual practice when employing mixed causal noncausal models, which consists in forecasting bubble bursts, we choose the stable period between 2017 and 2020 for the forecasting exercise. This allows us to forecast at farther horizons using a sampling importance resampling algorithm. We perform 1-, 3- and 6-months ahead probability forecasts. For short-horizons forecasts we correctly track probabilities to stay in the target bounds and interpret those results as the short-term credibility of the inflation targeting system based on historical inflation without the need to model people's beliefs. In longer horizon, probabilities seem to converge to a 50/50 result. We augment the univariate model with key economic variables and results show that during stable times, the addition of those variables do not impact significantly the predictions obtained from the univariate model. We then compare our measure of short-term credibility of the central bank with existing indices in the literature. We find that the distinct horizon and perspective used to build the indices makes our short-term credibility index complementary to the existing long-term indices. We employ receiver operating characteristic curves to determine the adequate probability threshold from which the Central bank is predicted to be credible or not.

\newpage
\section*{Appendix}
\appendix
\section{Simulations-based estimator}\label{bra-app:lanne}
Let $g(\varepsilon^*_+|u_T)$ be the conditional joint distribution of the \textit{M} future errors $(\varepsilon^*_{T+1},\dots,\varepsilon^*_{T+M})$, which using Bayes' Theorem can be expressed as follows, 
\begin{equation*}
    g(\varepsilon^*_+|u_T)=\frac{l(u_T|\varepsilon^*_+)}{l(u_T)} g(\varepsilon^*_+),
\end{equation*}
where \textit{g} is the \textit{pdf} associated with the error terms and \textit{l} denotes the densities associated with the process $u_t$. Thus, for any function $q$, the conditional expectation of such function of future error terms can be re-written as follows,
\begin{equation}
\begin{split}
    \mathbb{E}\Big[q(\varepsilon^*_+)\big|u_T\Big]&=\int q(\varepsilon^*_+)g(\varepsilon^*_+|u_T)d\varepsilon^*_+\\ &=\frac{1}{l(u_T)}\int q(\varepsilon^*_+)l(u_T|\varepsilon^*_+) g(\varepsilon^*_+)d\varepsilon^*_+\\
    &=\frac{\mathbb{E}_{\varepsilon^*_+}\Big[q(\varepsilon^*_+)l(u_T|\varepsilon^*_+)\Big]}{l(u_T)}.
    \end{split}
    \label{bra-eq:cond.exp.q}
\end{equation}

The conditional distribution $l(u_T|\varepsilon^*_+)$ can be obtained from the error distribution \textit{g}. Yet, since it is conditional on $\varepsilon^*_+$ instead of $u^*_{T+1}$, we can only obtain an approximation. Using this approximation and the Iterated Expectation theorem, the marginal distribution of $u_T$ can be approximated as follows, 
\begin{equation*}
    l(u_T)=\mathbb{E}_{\varepsilon^*_+}\big[l(u_T|\varepsilon^*_+)\big]\approx\mathbb{E}_{\varepsilon^*_+}\Bigg[g\Bigg(u_T-\sum_{i=1}^{M}\psi^i\varepsilon^*_{T+i}\Bigg)\Bigg].
\end{equation*}
Overall, by plugging the aforementioned approximation in \eqref{bra-eq:cond.exp.q}, we obtain
\begin{equation*}
    \mathbb{E}\Big[q(\varepsilon^*_+)\big|u_T\Big]\approx \frac{\mathbb{E}_{\varepsilon^*_+}\Bigg[q(\varepsilon^*_+)g\Big(u_T-\sum_{i=1}^{M}\psi^i\varepsilon^*_{T+i}\Big)\Bigg]}{\mathbb{E}_{\varepsilon^*_+}\Bigg[g\Big(u_T-\sum_{i=1}^{M}\psi^i\varepsilon^*_{T+i}\Big)\Bigg]}.
\end{equation*}

\section{Sample-based estimator for \textit{h}-step ahead forecasts}\label{bra-app:GJ}
The joint conditional predictive density \citep[as named by][]{filtering} or the causal transition distribution \citep[as named by][]{gourieroux2017local} of the \textit{h} future values of an \textit{MAR}(0,1) process, $(u^*_{T+1},\dots, u^*_{T+h})$, which is a Markov process of order one, can be obtained given the value of the last observed point $u_T$. While the interest is on predicting the future given contemporaneous and past information, it is only possible, by the model definition, to derive the density of a point conditional on its future point. Bayes' Theorem is applied repeatedly until all conditional \textit{pdf}'s are conditioned on the value of future points. The resulting predictive density of an \textit{h}-step ahead forecast is as follows,
\begin{equation} \label{bra-eq:ap:pred.dens2}
    \begin{split}
    & l(u^*_{T+1},\dots, u^*_{T+h}|u_{T}) \\
    &=l(u_{T},u^*_{T+1},\dots, u^*_{T+h-1}|u^*_{T+h})\times \frac{l(u^*_{T+h})}{l(u_{T})}\\
    &=\bigg\{l(u_T|u^*_{T+1},\dots,u^*_{T+h})l(u^*_{T+1}|u^*_{T+2},\dots,u^*_{T+h})\dots l(u^*_{T+h-1}|u^*_{T+h})\bigg \}\\
    &\hspace{1cm} \times \frac{l(u^*_{T+h})}{l(u_{T})},
    \end{split}
\end{equation}
where \textit{l} denotes densities associated with the noncausal process $u_t$. The process $u_t$ is a Markov process of order one, hence the conditioning information set of the densities can be reduced to a single future point. Furthermore, Equation \eqref{bra-eq:MAR.filter} states that $\varepsilon_t=u_t-\psi u_{t+1}$, thus, given the value of $u_{t+1}$, the conditional density of $u_t$ is equivalent to the density of $\varepsilon_t$ (which is known) evaluated at the point $u_t-\psi u_{t+1}$. Since $u_\tau=\psi u_{\tau+1}+\varepsilon_\tau$ and because $u_{\tau+1}$ and $\varepsilon_\tau$ are independent for all $1\leq \tau \leq T$, we have $f_{u_\tau|u_{\tau+1}}(x)=f_{\varepsilon_\tau+\psi u_{\tau+1}|u_{\tau+1}}(x)=f_{\varepsilon_\tau|u_{\tau+1}}(x-\psi u_{\tau+1})=f_\varepsilon(x-\psi u_{\tau+1})$. For simplicity, the density distributions related to $u_t$ (resp. $\varepsilon_t$) are just denoted by $l$ (resp. $g$). \\

In the absence of closed-form expressions for the marginal distributions $l(u^*_{T+h})$ and $l(u_{T})$ (when errors are Student's \textit{t} distributed for instance), the predictive density \eqref{bra-eq:ap:pred.dens2} can be approximated by substituting them with their sample counterparts using all past realised filtered values of $u_t$ \citep{filtering},

\begin{equation}
    \begin{split}
        &{l}(u^*_{T+1},\dots,u^*_{T+h}|\mathcal{F}_{T})\\ 
        & \approx {g}({u}_{T}-\psi u^*_{T+1})\dots {g}(u^*_{T+h-1}-\psi u^*_{T+h})
        \frac{N^{-1}\sum_{i=2}^{T}
        {g}(u^*_{T+h}-{\psi} u_i)}{N^{-1}\sum_{i=2}^{T} 
        {g}(u_{T}-{\psi} u_i)}.
    \end{split}
\end{equation}

\section{Results} \label{bra-sec:app.results}
\begin{table}[h!]\small
    	\centering
    	\caption{Probabilities to remain within the bounds at the forecasted points indicated based on different forecast horizon}
    	\begin{threeparttable}
    	\resizebox{\textwidth}{!}{%
        	\begin{tabular}{lc|ccccccccccc}
            \hline \hline 
                && \multicolumn{11}{c}{Dates} \\ \hline 
               && Dec-16 &Jan-17 &Fev-17 &Mar-17 &Avr-17 &May-17 &\cellcolor{gray!15}Jun-17 &\cellcolor{gray!15}Jul-17 &\cellcolor{gray!15}Aug-17 &\cellcolor{gray!15}Sep-17 &\cellcolor{gray!15}Oct-17 \\ \hline
               \multirow{2}{*}{1-month} 
                    & LLS &  0.540 &  0.659 &  0.991 &  0.996 &  0.997 &  0.991 & \cellcolor{gray!15}  0.877 & \cellcolor{gray!15}  0.201 & \cellcolor{gray!15}  0.143 & \cellcolor{gray!15}  0.079 & \cellcolor{gray!15}  0.156   \\
                    & SIR &  0.542 &  0.660 &  0.996 &  0.997 &  1.000 &  0.996 & \cellcolor{gray!15}  0.921 & \cellcolor{gray!15}  0.271 & \cellcolor{gray!15}  0.194 & \cellcolor{gray!15}  0.121 & \cellcolor{gray!15}  0.188  \\\hline
                \multirow{2}{*}{3-months} 
                    & LLS  &   &   &  0.491 &  0.694 &  0.887 &  0.879 & \cellcolor{gray!15}  0.894 & \cellcolor{gray!15}  0.771 & \cellcolor{gray!15}  0.612 & \cellcolor{gray!15}  0.351 & \cellcolor{gray!15}  0.369  \\
                    & SIR  &   &   &  0.498 &  0.720 &  0.916 &  0.914 & \cellcolor{gray!15}  0.930 & \cellcolor{gray!15}  0.862 & \cellcolor{gray!15}  0.721 & \cellcolor{gray!15}  0.471 & \cellcolor{gray!15}  0.482  \\ \hline
                \multirow{2}{*}{6-months} 
                    & LLS  &   &   &   &   &   &  0.500 & \cellcolor{gray!15}  0.580 & \cellcolor{gray!15}  0.663 & \cellcolor{gray!15}  0.664 & \cellcolor{gray!15}  0.671 & \cellcolor{gray!15}  0.628  \\
                    & SIR  &   &   &   &   &   &  0.553 & \cellcolor{gray!15}  0.626 & \cellcolor{gray!15}  0.717 & \cellcolor{gray!15}  0.742 & \cellcolor{gray!15}  0.759 & \cellcolor{gray!15}  0.720
  \\  \hline
                &\\
                &&\cellcolor{gray!15} Nov-17 &\cellcolor{gray!15}Dec-17 &\cellcolor{gray!15}Jan-18 &\cellcolor{gray!15}Fev-18 &\cellcolor{gray!15}Mar-18 &\cellcolor{gray!15}Avr-18 &\cellcolor{gray!15}May-18 &Jun-18 &Jul-18 &Aug-18 &Sep-18 \\ \hline
               \multirow{2}{*}{1-month} 
               & LLS & \cellcolor{gray!15}  0.295 & \cellcolor{gray!15}  0.358 & \cellcolor{gray!15}  0.590 & \cellcolor{gray!15}  0.307 & \cellcolor{gray!15}  0.323 & \cellcolor{gray!15}  0.151 & \cellcolor{gray!15}  0.301 &  0.426 &  0.976 &  0.996 &  0.996 \\
                    & SIR  & \cellcolor{gray!15}  0.322 & \cellcolor{gray!15}  0.401 & \cellcolor{gray!15}  0.620 & \cellcolor{gray!15}  0.354 & \cellcolor{gray!15}  0.356 & \cellcolor{gray!15}  0.168 & \cellcolor{gray!15}  0.321 &  0.447 &  0.990 &  0.999 &  0.999  \\ \hline
                \multirow{2}{*}{3-months} 
                    & LLS  & \cellcolor{gray!15}  0.307 & \cellcolor{gray!15}  0.448 & \cellcolor{gray!15}  0.547 & \cellcolor{gray!15}  0.560 & \cellcolor{gray!15}  0.640 & \cellcolor{gray!15}  0.488 & \cellcolor{gray!15}  0.510 &  0.393 &  0.523 &  0.574 &  0.411  \\  
                    & SIR  & \cellcolor{gray!15}  0.404 & \cellcolor{gray!15}  0.534 & \cellcolor{gray!15}  0.638 & \cellcolor{gray!15}  0.647 & \cellcolor{gray!15}  0.709 & \cellcolor{gray!15}  0.593 & \cellcolor{gray!15}  0.589 &  0.458 &  0.593 &  0.644 &  0.454  \\ \hline
                \multirow{2}{*}{6-months} 
                    & LLS  & \cellcolor{gray!15}  0.576 & \cellcolor{gray!15}  0.466 & \cellcolor{gray!15}  0.485 & \cellcolor{gray!15}  0.453 & \cellcolor{gray!15}  0.536 & \cellcolor{gray!15}  0.575 & \cellcolor{gray!15}  0.571 &  0.604 &  0.541 &  0.547 &  0.497  \\
                    & SIR  & \cellcolor{gray!15}  0.690 & \cellcolor{gray!15}  0.625 & \cellcolor{gray!15}  0.612 & \cellcolor{gray!15}  0.582 & \cellcolor{gray!15}  0.671 & \cellcolor{gray!15}  0.678 & \cellcolor{gray!15}  0.670 &  0.687 &  0.652 &  0.661 &  0.597  \\  \hline
                &\\
                && Oct-18 &Nov-18 &Dec-18 &Jan-19 &Fev-19 &Mar-19 &Avr-19 &May-19 &Jun-19 &Jul-19 &Aug-19 \\ \hline
               \multirow{2}{*}{1-month} 
                    & LLS  &  0.995 &  0.996 &  0.988 &  0.991 &  0.996 &  0.997 &  0.977 &  0.954 &  0.993 &  0.375 &  0.913   \\  
                    & SIR  &  0.999 &  1.000 &  0.996 &  0.997 &  1.000 &  0.999 &  0.982 &  0.964 &  0.998 &  0.540 &  0.943\\  \hline
                \multirow{2}{*}{3-months} 
                    & LLS  &  0.892 &  0.852 &  0.856 &  0.862 &  0.823 &  0.810 &  0.882 &  0.896 &  0.653 &  0.658 &  0.886 \\  
                    & SIR  &  0.931 &  0.897 &  0.880 &  0.881 &  0.891 &  0.876 &  0.917 &  0.935 &  0.693 &  0.696 &  0.915  \\   \hline
                \multirow{2}{*}{6-months} 
                    & LLS  &  0.560 &  0.578 &  0.344 &  0.643 &  0.669 &  0.589 &  0.643 &  0.645 &  0.647 &  0.674 &  0.679  \\  
                    & SIR  &  0.657 &  0.675 &  0.412 &  0.693 &  0.751 &  0.625 &  0.692 &  0.741 &  0.725 &  0.756 &  0.750  \\  \hline
                &\\
                && Sep-19 &\cellcolor{gray!15}Oct-19 &Nov-19 &Dec-19 &Jan-20 &  Fev-20 &  Mar-20 &  \cellcolor{gray!15}Avr-20 &  \cellcolor{gray!15}May-20 &  \cellcolor{gray!15}Jun-20 &  \cellcolor{gray!15}Jul-20 \\ \hline
               \multirow{2}{*}{1-month} 
                    & LLS  &  0.989 & \cellcolor{gray!15}  0.333 &  0.164 &  0.993 &  0.95 &     0.994 &      &  \cellcolor{gray!15}    &   \cellcolor{gray!15}   &   \cellcolor{gray!15}   &   \cellcolor{gray!15}  \\  
                    & SIR  &  0.996 & \cellcolor{gray!15}  0.415 &  0.201 &  0.998 &  0.961 &     0.998 &      & \cellcolor{gray!15}     &   \cellcolor{gray!15}   &   \cellcolor{gray!15}   &  \cellcolor{gray!15}   \\   \hline
                \multirow{2}{*}{3-months} 
                    & LLS&  0.306 & \cellcolor{gray!15}  0.695 &  0.857 &  0.410 &  0.453 &     0.882 &     0.492 &   \cellcolor{gray!15}  0.877 &  \cellcolor{gray!15}    &    \cellcolor{gray!15}  & \cellcolor{gray!15}   \\
                    & SIR &  0.454 & \cellcolor{gray!15}  0.778 &  0.904 &  0.510 &  0.579 &     0.910 &     0.527 &    \cellcolor{gray!15} 0.895 &   \cellcolor{gray!15}   &    \cellcolor{gray!15}  &   \cellcolor{gray!15}   \\ \hline
                \multirow{2}{*}{6-months} 
                    & LLS &  0.499 & \cellcolor{gray!15}  0.518 &  0.667 &  0.405 &  0.635 &     0.667 &     0.529 &    \cellcolor{gray!15} 0.514 &    \cellcolor{gray!15} 0.642 &     \cellcolor{gray!15}0.403 &   \cellcolor{gray!15}0.649\\
                    & SIR &  0.565 & \cellcolor{gray!15} 0.579 &  0.729 &  0.555 &  0.723 &     0.716 &     0.656 &     \cellcolor{gray!15} 0.665 &    \cellcolor{gray!15} 0.688 &     \cellcolor{gray!15}0.446 &     \cellcolor{gray!15}0.708\\ \hline
            \end{tabular}%
            }
        \end{threeparttable}
        \label{bra-tab:probas_in_bounds}
        \scriptsize \justifying Probabilities for inflation to remain between the bounds. LLS correspond to the simulations-based approach proposed by \citet{lanne2012optimal}, 1\,000\,000 were employed. SIR correspond to the sampling importance resampling algorithm based on the sample-based approach proposed by \citeauthor{filtering}, 100\,000 simulations were used in the first step, and 10\,000 in the resampling step. Shaded columns represent months during which inflation was below the bounds.
    \end{table}

\newpage
\section*{Acknowledgments}
The authors would like to thank Joann Jasiak and the participants of CFE 2021 for valuable suggestions. 

\bibliography{references.bib}

\end{document}